\documentclass[twocolumn,superscriptaddress,showpacs,prd,aps,amsmath,amssymb,nofootinbib]{revtex4-1}

\usepackage{graphicx}
\usepackage{longtable}

\newcommand{\beq}{\begin{equation}} 
\newcommand{\eeq}{\end{equation}} 
\newcommand{\beqn}{\begin{eqnarray}} 
\newcommand{\eeqn}{\end{eqnarray}}

\newcommand{\zD}{{\raise1.0ex\hbox{${}^{\ \circ}$}}\!\!\!\!\!D}
\newcommand{\alone}{{\raise0.5ex\hbox{${}^{\ 1}$}}\!\!\!\!\alpha}

\newcommand{\nalam}{\mathrel{\raise0.9ex\hbox{$^\lambda$}\mkern-14mu
\lower0.0ex\hbox{$\nabla$}}}

\newcommand{\zeroD}{{\raise1.0ex\hbox{${}^{\ \circ}$}}\!\!\!\!\!D}
\newcommand{\zLap}{{\raise1.0ex\hbox{${}^{\ \circ}$}}\!\!\!\!\Delta}
\newcommand{\zna}{{\raise1.0ex\hbox{${}^{\ \circ}$}}\!\!\!\!\!\nabla}
\newcommand{\zS}{{\raise1.0ex\hbox{${}^{\ \circ}$}}\!\!\!\!\!S}

\usepackage[normalem]{ulem}
\usepackage{epstopdf}
\usepackage{color}
\usepackage{soul}
\usepackage{ulem}
\usepackage{bm}
\usepackage{xspace}

\newcommand{\SACRA}{\textsc{sacra-mpi}\xspace}
\newcommand{\cocal}{\textsc{cocal}\xspace}

\usepackage[]{amsmath}
\usepackage[]{amsfonts}
\usepackage[]{amssymb}
\usepackage[]{mathrsfs}
\usepackage[]{mathtools}
\usepackage{times}
\usepackage{hyperref}
\hypersetup{
  colorlinks=true,        
  linkcolor=blue,         
  citecolor=cyan,         
}
\def\QEQ{{%
			\setbox0\hbox{$I$}%
			\rlap{\hbox to \wd0{\hss--\hss}}\box0
		}}

\begin{document}

\title{Evolution of bare quark stars in full general relativity: Single star case }

\author{Enping Zhou}
\affiliation{Max Planck Institute for Gravitational Physics (Albert Einstein
Institute), Am M\"uhlenberg 1, Potsdam-Golm, 14476, Germany} 

\author{Kenta Kiuchi}
\affiliation{Max Planck Institute for Gravitational Physics (Albert Einstein
	Institute), Am M\"uhlenberg 1, Potsdam-Golm, 14476, Germany} 
\affiliation{Center
	for Gravitational Physics, Yukawa Institute for Theoretical Physics, Kyoto
	University, Kyoto, 606-8502, Japan}
	
\author{Masaru Shibata}
\affiliation{Max Planck Institute for Gravitational Physics (Albert Einstein
	Institute), Am M\"uhlenberg 1, Potsdam-Golm, 14476, Germany} \affiliation{Center
	for Gravitational Physics, Yukawa Institute for Theoretical Physics, Kyoto
	University, Kyoto, 606-8502, Japan}

\author{Antonios Tsokaros}
\affiliation{Department of Physics, University of Illinois at Urbana-Champaign,
	Urbana, IL 61801, USA}

\author{K\=oji Ury\=u}
\affiliation{Department of Physics, University of the Ryukyus, Senbaru,
	Nishihara, Okinawa 903-0213, Japan}

\date{\today}

\begin{abstract}
We introduce our approaches, in particular the modifications of the primitive recovery procedure, to handle bare quark stars in numerical relativity simulations. Reliability and convergence of our implementation are demonstrated by evolving two triaxially rotating quark star models with different mass as well as a differentially rotating quark star model which has sufficiently large kinetic energy to be dynamically unstable. These simulations allow us to verify that our method is capable of resolving the evolution of the discontinuous surface of quark stars and possible mass ejection from them. The evolution of the triaxial deformation and the properties of the gravitational-wave emission from triaxially rotating quark stars have been also studied, together with the mass ejection of the differentially rotating case. It is found that supramassive quark stars are not likely to be ideal sources of continuous gravitational wave as the star recovers axisymmetry much faster than models with smaller mass and gravitational-wave amplitude decays rapidly in a timescale of $10\,$ms, although the instantaneous amplitude from more massive models is larger. As with the differentially rotating case, our result confirms that quark stars could experience non-axisymmetric instabilities
similar to the neutron star case but with quite small degree of differential rotation, which is expected according to previous initial data studies. 
\end{abstract}

\maketitle

\section{Introduction}
The detection of gravitational-wave (GW) signals from binary neutron star (BNS) merger events (GW170817 and GW190425)~\citep{Abbott2017, Abbott2020}, as well as the electromagnetic (EM) counterparts~\citep{Abbott2017b}, have provided significant
insights in the equation of state (EoS) of neutron stars (NSs). On the one hand, the inspiral GW signal puts a constraint on the tidal 
deformability of NSs~\citep{Abbott2018tidal,Abbottprx2019,Annala2017,De2018}, which translates into a constraint on the radius of NSs.
On the other hand, the EM counterparts indicate the fate of the merger remnant, which is tightly related to the maximum 
mass of cold and non-spinning NSs (or Tolman-Oppenheimer-Volkov maximum mass, $M_\mathrm{TOV}$) and the total mass of the binary. 
Various constraints have thus been put on the EoS of NSs accordingly 
\citep{Ruiz2017,Rezzolla2017,Shibata2019,Margalit2017,Bauswein2017b}.

Nevertheless, the state of matter at density as high as in NSs, particularly the role of quark matter, is still in open debate.
Recent studies have revealed that the existence of deconfined quark matter is possible in the high density core part of massive NSs, confronted with the most recent astrophysical observations \citep{Annala2020}. Many attempts aimed at searching for evidence of such a strong interaction phase transition (from conventional hadronic matter to quark matter) inside NSs from astrophysical observations have also been made~\citep{Bauswein2019b,Most2019}. As another possibility, 3-flavor quark matter could be the absolutely stable state of matter at high density~\citep{Bodmer1971,Witten84}. This may result in the existence of the 
so-called strange quark star (QS). Such object could exist above a certain NS mass (i.e., the two-family scenario~\citep{Bhattacharyya2017,Pietri2019}), and hence could be formed during the merger of BNSs. Alternatively, there are also suggestions that QSs could exist even for the normal NS mass range~\citep{Alcock86} and following this assumption the binary QS (BQS) merger scenario has also been 
studied in terms of the tidal deformability and mass ejections \citep{Zhou2018b,Bauswein2009,Bauswein2010}.

Figuring out the differences between the merger events involving QSs and NSs  
and verifying them with multi-messenger observations could greatly enrich our understanding in the nature of strong interaction. This requires
the modeling of QSs with the tool of general relativistic hydrodynamics as a first step. Previously, various efforts have been made in the calculation of the quasi-/equilibrium structures of
uniformly and differentially rotating QSs in general relativity (GR)~\citep{rosinska2000b,gourgoulhon1999,Stergioulas99a,Szkudlarek2019,Zhou2018,Zhou2019} as well
as the configuration of BQSs in the last orbits~\citep{limousin2005}. Despite all these 
attempts in constructing initial data for QSs, the progress in dynamical evolution of them is quite limited. In fact, there has been only one successful simulation of BQSs \citep{Bauswein2009,Bauswein2010}, in which only approximate GR and the smooth particle hydrodynamics method are adopted. It is both necessary and helpful for us to improve our understanding by pushing forward to evolve QSs in full GR grid-based hydrodynamics, which has yet been performed, yet.

The problem in directly evolving QSs results from the structural difference between QSs and NSs.
QSs are self-bound by strong interaction and the surface density is non-zero (hence also referred to as a 'bare' QS). Handling this discontinuity on the surface of QSs is a challenge in numerical simulation. In addition, the specific enthalpy of QSs
calculated in the conventional way does not go to unity as pressure goes to zero
\citep{Zhou2018}, and it may cause a problem when we recover fundamental thermodynamic variables from conserved variables which are directly evolved in numerical relativity (NR) (the so-called primitive recovery procedure).
Realizing this, we have tried to modify the NR code \SACRA \citep{Kiuchi2017b,Yamamoto2008} to resolve these two problems and achieved successful evolution of single QSs.
In this paper, we report the details of our
methods to handle QSs in NR and the first results of dynamical evolution of two triaxially rotating QS models and a differentially rotating QS model.

There are several motivations for us to tackle the problem of evolving triaxially and differentially rotating QSs as a first step. First, these systems provide us a good test-bed problem for evolving BQSs, for which we have to handle the motion of the sharp density discontinuity at the surface. The non-zero density surface of the triaxial QSs naturally moves in the computational domain during the evolution. This is also the case for differentially rotating stars with sufficiently large values of $T/|W|$ ratio (the ratio of kinetic energy, $T$, to gravitational potential energy, $W$) to be unstable for dynamical bar modes~\citep{Chandrasekhar1969book,Shibata:2000jt}. Thus, an essential technical element needed to handle the merger of BQSs can be tested by the simulation for these systems.
Secondly, previous studies indicated that triaxial deformation could play a more important role for QSs than for NSs, as
it could reach higher values of $T/|W|$~\citep{rosinska2003,Zhou2018}. Specifically, supramassive triaxial QSs (i.e., those with mass larger than $M_\mathrm{TOV}$) could exist for a large parameter space, while normally no such solution for NSs is present unless the EoS is extremely stiff in the crust and softer in the core~\citep{Uryu2016b,Zhou2018}. Therefore, a newly-born QS in
supernova or binary merger events may be a source for ground based GW detectors and is worthy to be investigated. Thirdly, determining the properties of mass ejection from binary merger events is a key for 
understanding the EM counterparts and we would like to make sure that our approach is capable of capturing the ejected mass. Mass ejection is expected from differentially rotating compact stars during the growth of the dynamical bar-mode instability~\citep{Shibata:2000jt} if $T/|W|$ is large enough. Thus we can confirm the ability of our methods by evolving such unstable differentially rotating configurations. 

The paper is organized as follows: In Sec.~\ref{sec:eos} we describe the EoS model used in this work and how we treat it in our numerical implementation. Details about the setup of our initial data solver and dynamical evolution code as well as the convergence tests are presented in Sec.~\ref{sec:numerical}. 
The results for the evolution of the QS models, as well as the physical interpretation of the results are reported in Sec.~\ref{sec:results}. In the end, we briefly conclude and discuss future prospects of evolving QSs in Sec.~\ref{sec:disandconclu}. More quantitative tests on the code performance are found in the Appendix. Throughout this paper, $c$ and $G$ denote the speed of light and the gravitational constant, respectively.

\section{Quark Star Equation of State}
\label{sec:eos}

The most widely used EoS for QSs is the MIT bag model, which is initially used as a phenomenological model for hadrons \citep{chodos1974}. Modern versions of the model
take into account more nuclear physics details, such as perturbative quantum choromodynamics (QCD)
corrections due to the gluon-mediated interaction between quarks as well as the finite mass of 
strange quarks \citep{Fraga2001,Alford2005}. As our first attempt to evolve QSs, 
we neglect these details and stick with the simplest form of the MIT bag model, which could be 
written as:
\beq
p={1\over3}(e-4B),
\label{eq:mit}
\eeq 
where $p$ is the pressure for given energy density $e$, and $B$ is the bag constant which is 
related to the finite surface density. Ideally, the pressure of quark matter vanishes at $e=4B$ and density drops to 0 discontinuously across the surface; that is, Eq.~(\ref{eq:mit}) is valid only for $e \geq 4B$. However, matter with density lower than the surface density exists in nature. Even for the case of 'bare' QS (which is exactly the model considered in this paper) that is not supposed to possess a low density crust consisting of conventional hadronic matter, such low density matter is still relevant when mass ejection is present (for instance, in the cases of dynamically unstable models and BQS mergers). Theoretically, the exact form and fate of the material ejected from a BQS merger are unclear. Some previous studies suggested that the ejection from QSs is likely to be in the form of strange quark nuggets~\citep{Madsen1988,Caldwell1991}, which are also self-bound as the bare QS itself (i.e., the pressure is also zero on the surface of the ejected nuggets). The main source of pressure from the low density ejecta of QSs is therefore the thermal contribution. Nevertheless, more recent studies indicate that the possibility of ejected quark matter decaying into normal nucleon matter does exist \citep{Paulucci2017,Bucciantini2019}. In this paper, we will not consider more complicated nuclear physical process at the moment and model such matter as ideal gas (the details of which is found in Sec.~\ref{sec:nm}).

To obtain the rest-mass density and specific enthalpy which are variables necessary for 
relativistic hydrodynamics, we need the 
more explicit form of the MIT bag model as
\beq
\begin{split}
p=K\rho^{4/3}-B, \\
e=3K\rho^{4/3}+B,
\end{split}
\label{eq:mitexplicit}
\eeq
where $\rho$ could be regarded as the rest-mass density~\footnote{A more popular version of this explicit form is written in terms of the number density $n$ rather than the rest-mass density $\rho$. We can always rewrite it by assuming certain rest mass for quarks. For example, 931\,MeV/$c^2$ is usually assumed for 3 quarks~\citep{Bhattacharyya2016}.} and $K$ is constant. The choice of $K$ is usually determined by details of interactions between quarks in the system \citep{Fraga2001,Alford2005,Weissenborn:2011qu,Bhattacharyya2016}, namely $K\sim a_4^{-1/3}$ in which $a_4$ is a model parameter that characterizes the perturbative QCD corrections. However, as discussed in many previous studies~\citep{Bhattacharyya2016,Li2017,Zhou2018}, the choice of $a_4$ (hence $K$) has only negligible influence on the EoS model once the bag constant is fixed. In particular, in the case that we neglect the mass of strange quarks (which is exactly the case assumed in this work), the choice of $K$ has no impact on the EoS model at all. This could be 
easily understood if one eliminates $\rho$ from Eq.~\eqref{eq:mitexplicit} to recover Eq.~\eqref{eq:mit}, regardless of whatever value for $K$ is used. Consequently, we could 
employ a value of $K$ which is most convenient for us to implement numerically, and $\rho$ 
becomes a purely auxiliary variable \footnote{In this sense baryonic mass of the star could alter with different choice of $K$. However, it does not affect any other quantities and hydrodynamics~\citep{Bhattacharyya2016}.}.

With the expression of Eq.~\eqref{eq:mitexplicit}, the specific enthalpy is written as
\beq
h=\frac{e+p}{\rho}=4K\rho^{1/3}.
\label{eq:mith}
\eeq
On the surface of the QSs, the density $\rho_\mathrm{s}$ is determined by the fact that the pressure $p_\mathrm{s}$ should vanish:
\beq
p_\mathrm{s}=K\rho_\mathrm{s}^{4/3}-B=0,
\eeq
which reads
\beq
\rho_\mathrm{s}=\left({B \over K}\right)^{3/4}.
\eeq
Hence the specific enthalpy on the surface of QSs in this model is 
\beq
h_\mathrm{s}=4K\rho_\mathrm{s}^{1/3}=4K^{3/4}B^{1/4}.
\eeq
Now it becomes obvious that normally with a choice of $K$ that is motivated by nuclear physics, $h_\mathrm{s}$ does not approach $c^2$. This could bring another source of discontinuity, because for the case of the normal matter for which the internal energy is much smaller than the rest-mass energy, the specific enthalpy approaches $c^2$. In particular, most of the physically motivated choice of $K$ for the quark matter results 
in $h_\mathrm{s}<c^2$. This implies that the specific enthalpy is not only discontinuous across the surface but also not monotonic, and $\rho$ as a function of $h$ is not even a single-valued function. This could make the primitive recovery procedure extremely complicated, if ever possible. 

\begin{figure}
	\begin{center}
		\includegraphics[height=70mm]{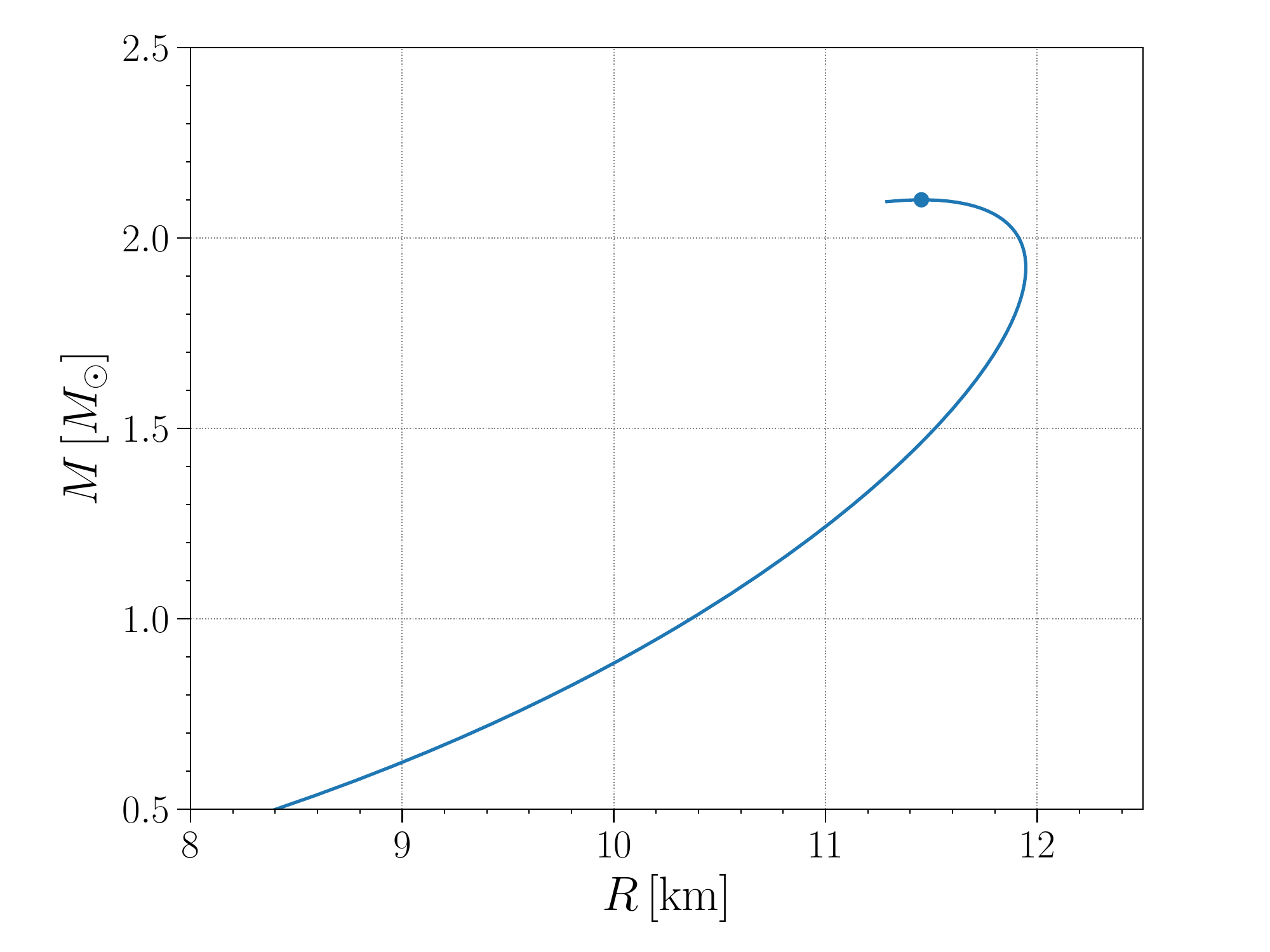}
	\end{center}
	\caption{Mass radius relation for cold non-rotating QSs with the MIT bag model adopted in this paper. Labeled by the filled circle is the maximum mass solution, with $M_\mathrm{TOV}=2.100\,M_\odot$ and $R_\mathrm{TOV}=11.5\,$km.}
	\label{fig:tov}
\end{figure}

Taking into account the fact that changing $K$ does not matter much to the EoS modeling for our purpose, it is then quite straightforward to make a choice for $K$ such that $h_\mathrm{s}=c^2$, which requires
\beq
K=\left(\frac{c^8}{256B}\right)^{1/3}.
\label{eq:mitk}
\eeq
This implies that for every choice of the bag constant, we could determine a unique choice for Eq.~\eqref{eq:mitexplicit} that resolves the issue of an ill-behaved function of $h(\rho)$. The EoS parameters
adopted in this work and also properties of the spherical QSs are listed in Table~\ref{tab:eospar} and the mass-radius relation is shown in Fig.~\ref{fig:tov}.

\begin{table}
	\begin{tabular}{cccccccccccc}
		\hline
		$B$\,[$\mathrm{erg\,cm^{-3}}]$ & 8.3989$\times10^{34}$ \\ $K$\,[$\mathrm{g^{-1/3}cm^{3}s^2}]$ &  3.1191$\times10^{15}$\\
		$\rho_\mathrm{s}\,[\mathrm{g\,cm^{-3}}]$ & 3.737$\times10^{14}$\\
		$M_\mathrm{TOV}\,[M_\odot]$ & 2.100 \\
		 $R_\mathrm{TOV}\,[\mathrm{km}]$ & 11.5 \\
		 $R_{1.4}\,[\mathrm{km}]$ & 11.3 \\
		  $\Lambda_{1.4}$ & 598 \\
		\hline
	\end{tabular}
	\caption{Parameters for the MIT bag model adopted in this paper. $B$ and $K$ are the bag constant and coefficient as in Eq.~\eqref{eq:mitexplicit}. $\rho_\mathrm{s}$ is the rest-mass density on the surface of the QS. The numbers listed for the EoS parameter are the values in cgs units. Mass and radius of the maximum-mass spherical solution as well
	as radius and tidal deformability of the $1.4M_\odot$ solutions are also listed. This model
    is consistent with the observations of the most massive pulsars \citep{Antoniadis2013,Cromartie2020} as well as the upper limit of tidal deformability \citep{Abbott2017}.}
	\label{tab:eospar}
\end{table}

Although shock heating does not play an important role for the evolution of a 
single star, we still employ the $\Gamma_\mathrm{th}$ prescription to take into account finite temperature effects as we ultimately will try to deal with BQS systems. In this model, the pressure is expressed as a sum of two components, the cold part ($p_\mathrm{cold}(\rho)$) determined by Eq.~\eqref{eq:mitexplicit} and a thermal part related to the specific internal energy density as
\beq
p=p_\mathrm{cold}(\rho)+(\Gamma_\mathrm{th}-1)\rho\epsilon_\mathrm{th},
\label{eq:ptotal}
\eeq
where $\epsilon_\mathrm{th}$ is the thermal part of specific internal energy density defined by
\beq
\epsilon_\mathrm{th}=\epsilon-\epsilon_\mathrm{cold}(\rho).
\eeq
Here we follow the normal definition such that specific internal energy density $\epsilon=e/\rho-c^2$. It is worth noting that only with the choice of $K$ according to Eq.~\eqref{eq:mitk}, are we able to guarantee that $\epsilon_\mathrm{cold}$ approaches zero on the surface of the star.

In this work, $\Gamma_\mathrm{th}$ is chosen to be $4/3$ which makes the primitive recovery process much easier to do for QSs (which we will demonstrate later in Sec.~\ref{sec:nm}), although in principle any other values are 
also possible to be employed at a higher computational cost. 

\section{Numerical Setup}
\label{sec:numerical}
\subsection{Initial Data}
\label{sec:id}
The quasi-equilibrium configurations of rotating QSs are calculated by the 
initial data solver \cocal (Compact Object CALculator) \citep{Tsokaros2015,Uryu2012}.
The implementation of QS EoS models as well as convergence and accuracy tests are found in our previous papers for axisymmetrically uniformly/differentially rotating and triaxially uniformly rotating cases~\citep{Zhou2018,Zhou2019}.

For the triaxial rotation case, as mentioned before, one significant difference between NSs and QSs is that supramassive triaxially rotating QSs could exist for a vast space of parameters. To explore both the supramassive case (which could be formed during a binary merger event) as well as a relatively standard mass case (which could be formed in a supernova), we have built two solution sequences with different values of the central density. The way of constructing the solution sequences is the same as in~\citep{Zhou2018}: We construct the sequence of solutions for a fixed value of the central density with decreasing 
$R_z/R_x$ ratio without imposing axisymmetry during the calculation. Here the direction of $z$-axis corresponds to the angular momentum direction of the rotating star and $R_z$ refers to the coordinate radii of the star along $z$-axis. Since we are considering triaxial configuration here, $R_x$ refers to the coordinate radii of the longest axis in the equatorial plane and $R_y$ to the shortest. As a consequence, in the presence of the solution with $R_z/R_x$ small enough to possess a sufficiently high value of the $T/|W|$ ratio for bifurcating into
the triaxial solution branch, a star with $R_y/R_x<1$ is spontaneously obtained. We then
pick up a triaxial solution with mass that is astrophysically interesting and use it as our
initial data.

As with the differential rotation case, the widely used $j$-const law
\citep{Komatsu89,Ansorg2009,Rosinska2017,Studzinska2016,Baumgarte00b,Morrison2004,Kaplan2014}:
\beq
j(\Omega)=A^2(\Omega_\mathrm{c}-\Omega), 
\label{eq:jconst} 
\eeq 
is chosen to construct the solutions, in which $A$ is a parameter that characterizes the degree of differential rotation and a normalized version $\hat{A}=A/r_e$ is also often used where $r_e$ is the equatorial coordinate radius of the star (for smaller values of $\hat{A}$, the differential rotation degree is higher). Previous initial-data studies~\citep{Zhou2019,Szkudlarek2019} show that the properties of differentially rotating QSs for a given value of $\hat{A}$ is quite different from that of NSs. In particular, continuous transition to toroidal sequences (i.e., type C solution according to the classification of~\citep{Ansorg2009}) happens at lower degree of differential rotation (at $\hat{A}\sim 3$ for QSs while at $\hat{A}=1$ for NSs). In previous studies for NSs~\citep{Shibata:2000jt,Baiotti06b}, the $\hat{A}$ parameter is typically chosen to be unity or smaller for the case of NSs to explore the dynamical bar-mode instability. Taking the difference between QSs and NSs mentioned above into account, we choose $\hat{A}=3.0$ for the MIT bag model and pick up a solution with $T/|W|$ large enough as the initial data. In addition, we evolve a differentially rotating NS model with the APR4 EoS~\citep{Akmal1998a} and $\hat{A}=1.0$ for comparison purpose.

The quantities estimated for the initial-data models we consider in this work are listed in Table~\ref{tab:id}. In the following, the triaxial model with lower mass will be referred to as MIT148 and the supramassive triaxial model as MIT265, according to their ADM mass. The differential rotation models will be referred to as MIT275dr and APR206dr for the QS and NS cases, respectively.

\begin{table*}
	\begin{tabular}{ccccccccccccc}
		\hline
	    Model & $R_x\,$[km] & $R_z/R_x$ & $R_y/R_x$ & $\rho_m$\,$[\mathrm{g\,cm^{-3}}]$ &  $M_\mathrm{ADM}\,[M_\odot]$ & $M_\mathrm{b}\,[M_\odot]$ & $J\,$[erg\,s] & $T/|W|$ & $P_\mathrm{c}\,[\mathrm{ms}]$ & $\hat{A}$ & $\dot{E}\,[\mathrm{erg\,s^{-1}}]$ &
		\\
		\hline
		MIT148    & 14.0 (18.3) & 0.457 (0.469) & 0.719 (0.728)  & $5.0\times10^{14}$ & 1.488 & 1.636 & $1.81\times10^{49}$ & 0.1712 & 0.955 & - & $3.51\times10^{53}$ &	 \\
		MIT265  & 13.5 (21.9) & 0.458 (0.486) & 0.781 (0.802) & $7.6\times10^{14}$ &
		2.655  & 3.109 & $5.23\times10^{49}$ & 0.1872 & 0.774 & - & $8.50\times10^{53}$ &  \\
		MIT275dr & 14.4 (22.3) &   0.313 (0.327) & - &   $5.0\times10^{14}$ &  
		2.757 &  3.121 & $7.35\times10^{49}$ & 0.2867 & 0.705 & 3.0 & - & \\
		APR206dr & 13.8 (19.6) & 0.250 (0.260) & - & $6.0\times10^{14}$ & 2.060 & 2.251 & $3.92\times10^{49}$ & 0.2599 &
		0.409 & 1.0 & - & \\
		\hline
	\end{tabular}
	\caption{Quantities of three single QS models and one NS model 
     	considered in this work, as estimated according
		to the initial data solutions. $R_x$ is the coordinate (proper) equatorial radius 
		and $R_z/R_x$ is the ratio of coordinate radius along $z$- and $x$-axis and $R_y/R_x$ is for $y$- to $x$-axis ratio. $\rho_m$ is the
		maximum rest-mass density inside the star. $M_{\rm ADM}$, $M_{\rm b}$, $J$, $T/|W|$ , $P_\mathrm{c}$ and $\hat{A}$ are Arnowit-Deset-Misner mass, baryonic mass, angular momentum, ratio between kinetic and gravitational potential energy, the central rotation period, and the differential rotation parameter in the $j$-const law.  Definitions can be found in the Appendix of~\citep{Uryu2016a}. $\dot{E}$ is the luminosity of the GW at $t=0$ as estimated by the quadrupole formula which is a good approximation for the instantaneous GW strain in the beginning of the dynamical evolution.} 
	\label{tab:id}
\end{table*}

\subsection{Dynamical Evolution}
\label{sec:nm}
The dynamical evolution of the initial data is performed with our numerical relativity code \SACRA. To solve Einstein's evolution equation, a moving puncture version of the Baumgarte-Shapiro-Shibata-Nakamura formalism~\citep{Shibata95,Baumgarte99,Campanelli06,Baker:2005vv} is employed in the code. A constraint propagation prescription similar to Z4c scheme~\citep{Hilditch2012} is also implemented in part of the computational domain (cf.~\citep{Kyutoku2014} for details). More details about \SACRA, such as the finite differencing schemes as well as the adaptive/fixed mesh refinement (AMR/FMR) setups, are found in~\citep{Yamamoto2008}.

In numerical hydrodynamics, we employ a high-resolution shock capturing scheme. For this case, one always has to impose a small but non-zero density for the exterior of the star to recover the 4-velocity (which is a primitive variable) from the momentum density (which is an evolved variable).
In this work, this  density is set to be  $10^{-12}$ times of the surface density of our QS model \footnote{Note that \cocal calculates QS initial data with finite surface density. Consequently, the floor density of $10^{-12}\rho_\mathrm{s}$ has to be imposed in the beginning of the evolution. An initial tiny specific internal energy density of $4K\rho_{\mathrm{floor}}^{1/3}$ is assigned to the floor density, to be consistent with Eq.~(\ref{eq:mith}) in the $p=0$ limit.}. Nevertheless, the situation is a bit more complicated than NS cases, even if we manage to choose a model according to Eq.~\eqref{eq:mitk} such that the specific enthalpy is now well-behaved across the surface of QSs. For NS cases, Eq.~\eqref{eq:ptotal} is valid in the entire computational domain, with $p_\mathrm{cold}$ calculated with exactly the same EoS both for the NS and the region outside the NS. However, it is obvious that this is not the case for QSs, because $p_\mathrm{cold}$ vanishes for $\rho<\rho_\mathrm{s}$. As a consequence, only the thermal component in Eq.~(\ref{eq:ptotal}) exists for the region outside the QSs:
\beq
p=(\Gamma_\mathrm{th}-1)\rho\epsilon,
\label{eq:atmeos}
\eeq 
and we choose $\Gamma_\mathrm{th}=4/3$ for $\rho<\rho_{\mathrm{s}}$, same as the thermal contribution of the matter in the QSs such that the EoS model is consistent throughout the inside and outside of the star. 

Having to use two different EoS models for the QS and the other brings another problem. In relativistic hydrodynamics, the conserved variables are evolved in every time step, and
the primitive variables (such as the rest-mass density, 4-velocity, and specific internal energy) have to be determined from the conserved variables with the help of EoS and normalization relation of the 4-velocity (i.e., by the so-called primitive recovery procedure). However,
since we have no idea about the value of the rest-mass density before the primitive recovery, we do not know whether the part of matter whose primitive variables to be recovered is inside the QS or not. 
As a result, we do not know which EoS model is adopted for the primitive recovery procedure in advance. Therefore, we have to modify the primitive recovery part to incorporate QS EoS models. We will briefly explain our strategy in the following.

For our implementation of relativistic hydrodynamics, the conserved variables which are directly evolved are
\beq
\rho_*=\rho w\sqrt{\gamma},
\label{eq:rhostar}
\eeq
\beq
S_i=\rho_*hu_i,
\label{eq:si}
\eeq
\beq
S_0=\rho_* \left(hw-\frac{p}{\rho w}\right),
\label{eq:s0}
\eeq
where $\sqrt{\gamma}$ is the determinant of the 3-metric on the spacelike hypersurface, $\gamma_{ij}$, and is solved during the time evolution, $u_i$ is the spatial component of the 4-velocity of the fluid element, $p$, $\rho$, $h$ and $w$ are the total pressure (including the thermal contribution), rest-mass density, specific enthalpy, and Lorentz factor of the fluid, which we would like to recover. Including the three components in $u_i$, we have in total 7 unknown variables and only 5 relations between them (Eqs.~\eqref{eq:rhostar}--\eqref{eq:s0}), and hence, additional two relations are needed. For this, one is the EoS model and the other is the normalization relation of the 4-velocity, which is rewritten to the definition of Lorentz factor as
\beq
w^2=1+\gamma^{ij}u_iu_j=1+\gamma^{ij}\frac{S_iS_j}{\rho^2_*h^2},
\label{eq:pr1}
\eeq
where Eq.~\eqref{eq:si} was used for the second equality. We here define $q^2 \equiv \gamma^{ij}S_iS_j/\rho^2_*$ (which is purely determined during the evolution) in the following to simplify the expressions. Another given quantity is defined using Eqs.~\eqref{eq:rhostar} and~\eqref{eq:s0} as
\beq
e_0=\frac{S_0}{\rho_*}=hw-\frac{p\sqrt{\gamma}}{\rho_*}.
\label{eq:pr2}
\eeq
Note that EoS provides us $p$ as a function of $w$ (since $\rho_*$ and $\sqrt{\gamma}$ are obtained during evolution, $\rho$ can be obtained once $w$ is determined according to Eq.~\eqref{eq:rhostar}) and $h$. Therefore, Eqs.~\eqref{eq:pr1} and~\eqref{eq:pr2} are two equations for two unknowns $h$ and $w$. Thus, the primitive recovery, in essence, is to find a solution for $h$ and $w$ using the above two algebraic equations.

For the region outside the QSs, for which the EoS is Eq.~\eqref{eq:atmeos}, the following algebraic equation for $h$ is obtained by substituting out $w$ from Eqs.~\eqref{eq:pr1} and~\eqref{eq:pr2}:
\beq
e_0^2(h^2+q^2)-\Gamma_\mathrm{th}^{-2}[h^2+(\Gamma_\mathrm{th}-1)h+\Gamma_\mathrm{th}q^2]^2=0,
\label{eq:atmpr}
\eeq
which we can solve by root finding methods. For the QS with 
$\Gamma_\mathrm{th}=4/3$, we have realized that the equation can be greatly simplified and Eq.~\eqref{eq:pr2} becomes
\beq
e_0=h\left(w-\frac{1}{4w}\right)+\frac{B\sqrt{\gamma}}{\rho_*},
\eeq
with which we obtain the following equation for $w$:
\beq
(s^2-q^2)w^4+\left(\frac{q^2}{2}-s^2\right)w^2-\frac{q^2}{16}=0.
\eeq
Here, we defined $s=e_0-B\sqrt{\gamma}/\rho_*$. A solution can be obtained analytically \footnote{In the case of other choices of $\Gamma_\mathrm{th}$, terms of other powers of $w$ are involved in the equation and no analytical solution can in general be obtained. Instead we have to do a root finding for more general choices of $\Gamma_\mathrm{th}$.} for $w^2$ as long as $s^4-(3/4)q^2s^2$ is larger than 0. We can then take the root for $w^2$ which is larger than 1, if exists, and recover the value for $\rho$ to see whether it is indeed larger than $\rho_\mathrm{s}$. In the case that the recovered value of $\rho$ is actually smaller than $\rho_\mathrm{s}$ or there exists no root for $w^2$ that satisfies $w\geq 1$, we treat the fluid as the non-quark matter, and solve Eq.~\eqref{eq:atmpr} with the Newton-Raphson method to find a solution. We succeeded in evolving the two triaxially rotating and one differentially rotating QS models with this implementation. In particular, it is worth mentioning that for the case of triaxial solutions and dynamically unstable differentially rotating solution, the surface of a QS is moving in the computational domain. This implies that there exists certain grid points on which the fluid is sometimes inside the QS and sometimes outside. The fact that the triaxial solutions can be evolved with this method verifies its reliability. We will focus on the tests of the code performance in the next subsection.

\subsection{Code Tests}
\label{sec:test}

We employ three different grid setups in this work to explore the accuracy and convergence behavior of the code in evolving QSs.  Nine FMR levels, which are all centered at the QS core, are constructed for all the three grid configurations, with every level doubling the size and grid intervals of its finer level.
In every refinement level, uniform and vertex-centered Cartesian coordinates are used to cover
the $x, y$, and $z$ directions with $2N+1,2N+1$ and $N+1$ points (equatorial plane symmetry is assumed) and we choose $N=80, 120$, and 160 for three different setups. The distance from the coordinate origin to the boundary of the finest level along each axis is set
to be 16\,$GM_\odot/c^2$ ($\approx 23.6\,$km) for all the three setups; i.e., an outer boundary of the entire FMR domain is as large as $16 \times 2^8=4096\,GM_\odot/c^2\approx 6048\,$km. The boundary is much larger than the typical GW wavelength of the models we considered here,  $\lambda_\mathrm{gw}\sim\pi/\Omega$, which is of the order of 100\,$GM_\odot/c^2\approx 147.7\,$km according to the rotational period found from the initial data. It is straightforward to find the grid resolution in the finest level as $\Delta_x=\Delta_y=\Delta_z=295,~197$, and $148\,$m for three different resolutions.

\begin{figure}
	\begin{center}
		\includegraphics[height=70mm]{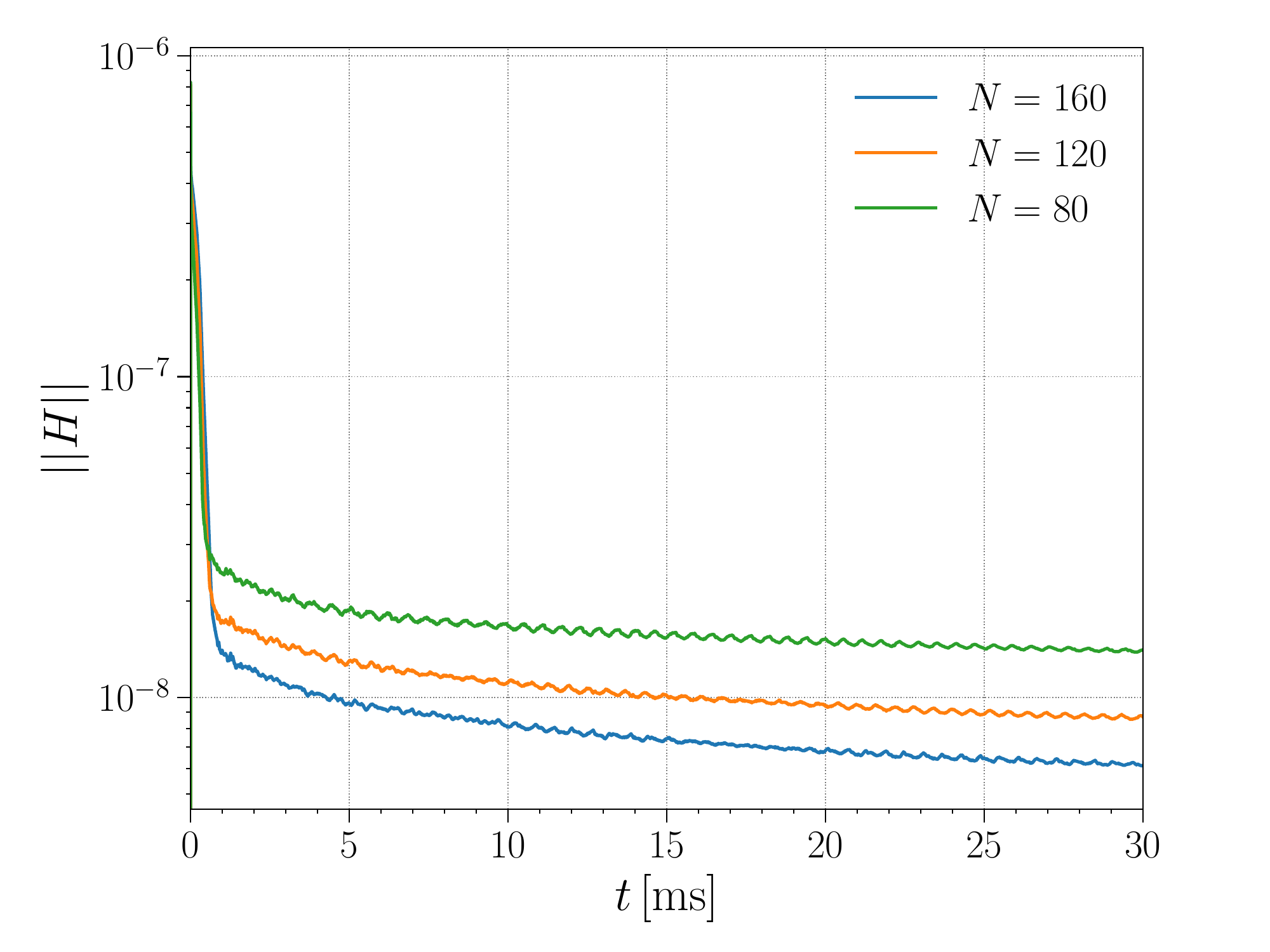}
		\includegraphics[height=70mm]{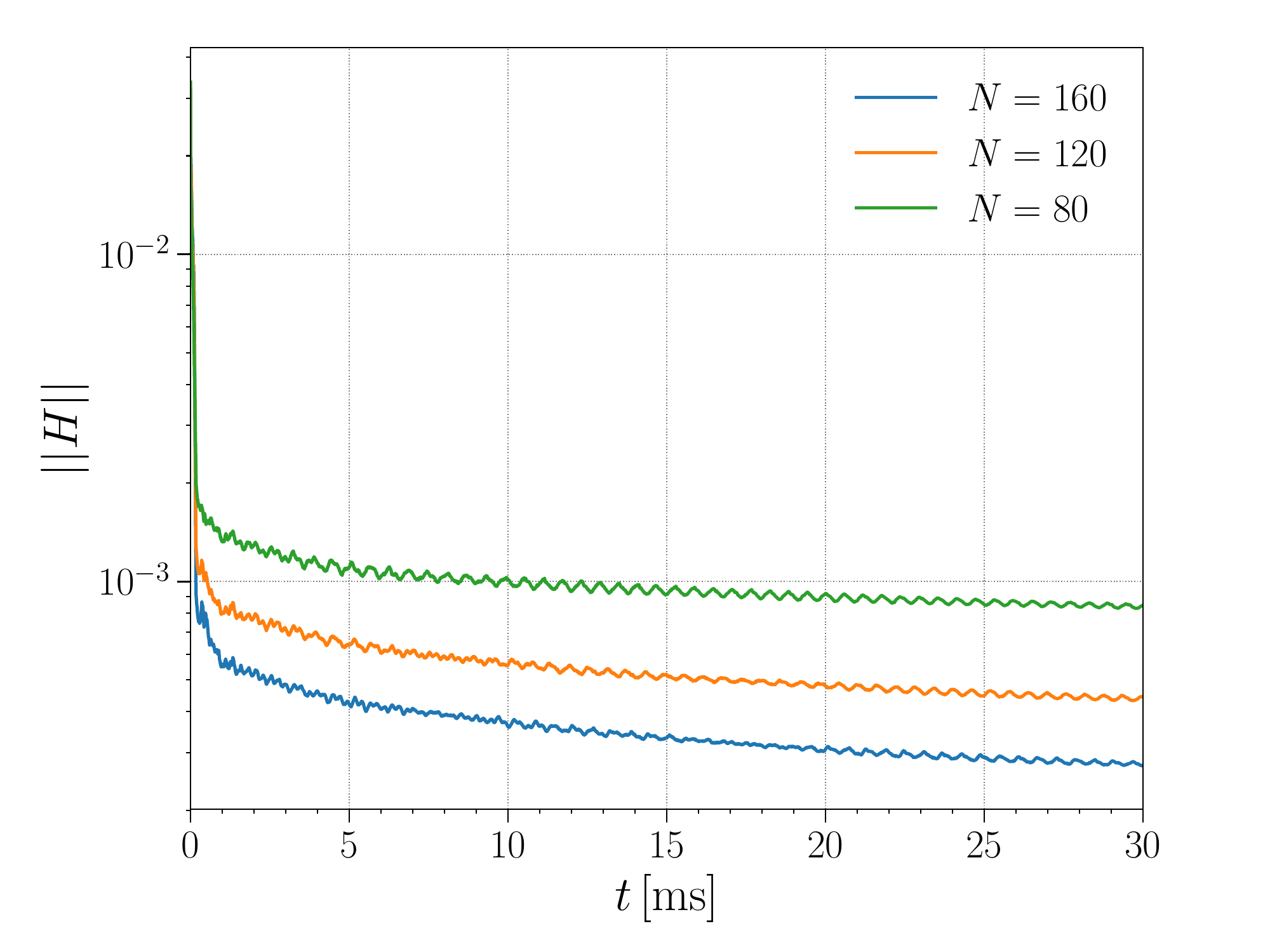}
	\end{center}
	\caption{The Hamiltonian constraint violation for model MIT148 during the evolution. The upper panel shows the volume-averaged result and the lower one for density-weighted result. Results of resolutions from low to high correspond to the curves from top to bottom (green for $N=80$, orange for $N=120$, and blue for $N=160$).}
	\label{fig:cv148}
\end{figure}

\begin{figure}
	\begin{center}
		\includegraphics[height=70mm]{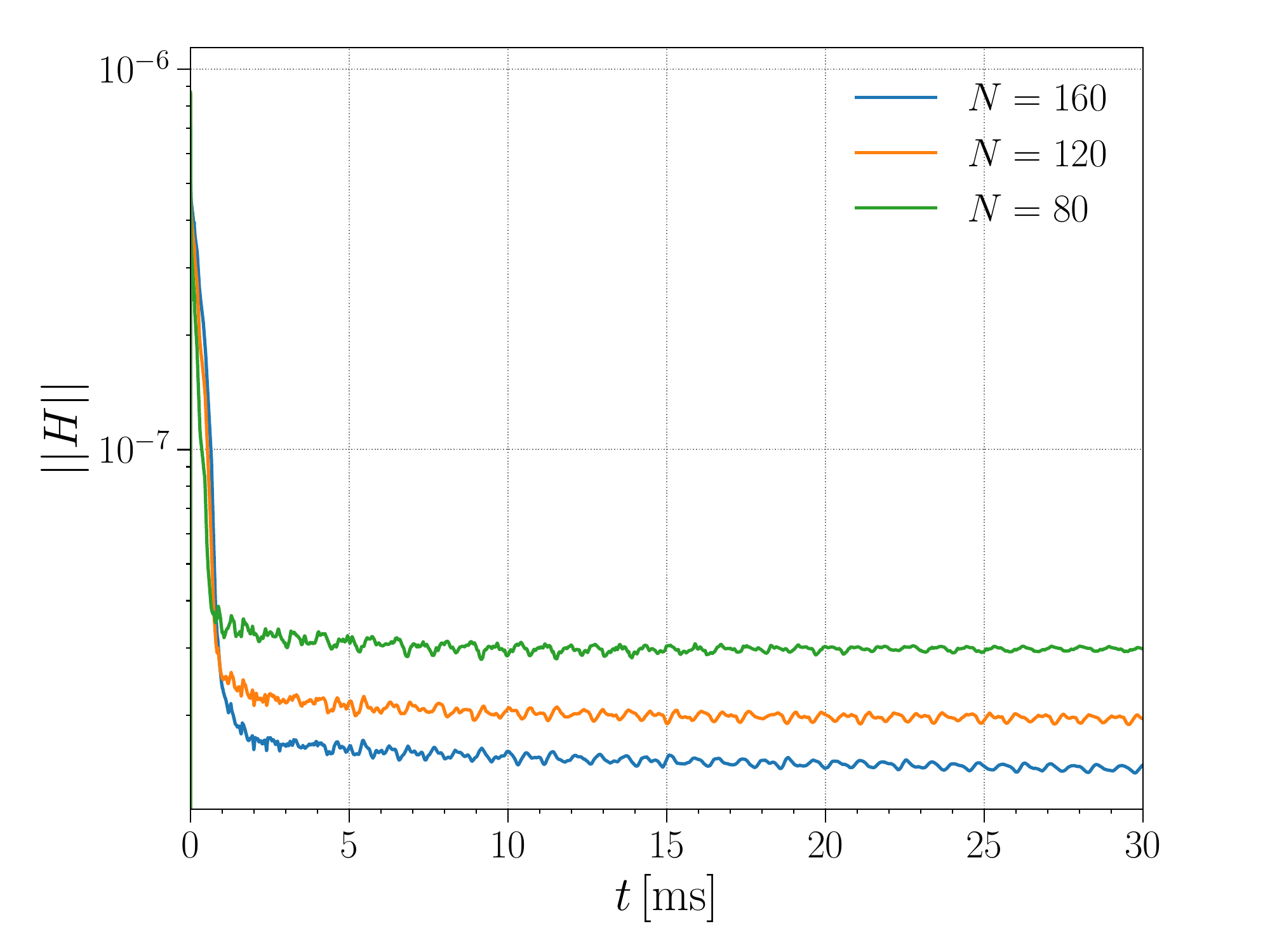}
		\includegraphics[height=70mm]{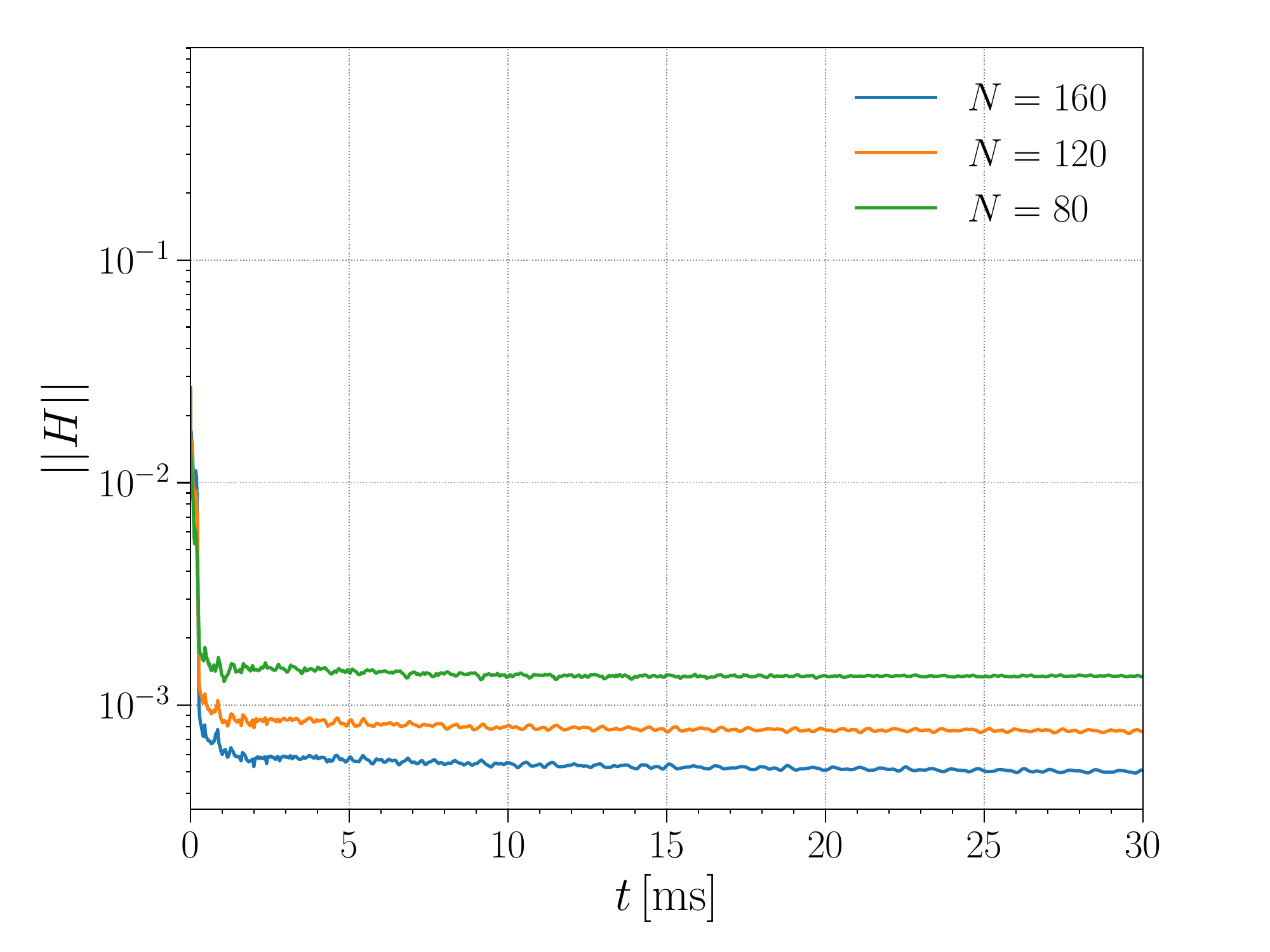}
	\end{center}
	\caption{Same as Fig.~\ref{fig:cv148} but for model MIT265.}
	\label{fig:cv265}
\end{figure}

We monitor both the volume-averaged and density-weighted Hamiltonian constraint violations for all the three resolutions up to $t=30$\,ms and show them in 
Figs.~\ref{fig:cv148} and~\ref{fig:cv265} for models MIT148 and MIT265, respectively. 
The initial constraint violation, which is mainly a result of the interpolation of the initial data from the surface fit coordinate employed in \cocal to the Cartesian coordinates in \SACRA, is significantly reduced at the early stage of the dynamical evolution, due to the constraint violation propagation scheme. After that, the constraint violation stays 
at a relatively low level for the entire duration of the run. 

The convergence behavior is also studied by analyzing this result. A certain quantity evaluated at one grid resolution $f_N$ could be expressed as 
\beq
f_N=f_\mathrm{exact}+A_f\Delta_N^{\zeta}, 
\label{eq:convergence}
\eeq
where $f_\mathrm{exact}$ is the exact solution if one has infinitely fine resolution, $\Delta_N$ is the grid interval associated with the resolution $N$, $A_f$ is a constant, and $\zeta$ is the convergence order of the code. For constraint violations, one should expect zero violation at infinite resolution. Hence, we could directly divide the constraint violation value at a given time in different resolutions and figure out $\zeta$, given the ratio of grid intervals between two resolutions. By doing so, we have found approximately linear order convergence ($\zeta \approx 1.0$) for volume-averaged constraint violation for any given time in the simulation. This agrees with our expectation: On one hand, we have already shown in our previous study that in the case of QSs, many quantities obey first-order convergence~\citep{Zhou2018} for constructing initial data due to the presence of the density discontinuity at the QS surface; on the other hand, as a minimum modulus flux-limiter function (for details,  cf.~\citep{Shibata_book:2016}) is used in \SACRA, the order of the interpolation accuracy during reconstruction could decrease to approximately linear order if we encounter strong discontinuity to avoid unphysical oscillations: This is exactly the case on the surface of a QS. The density-weighted result shows a convergence order of $\approx 1.4$ for MIT265 and 1.5 for MIT148, which further supports this understanding: The inner part of the QS, for which no discontinuity is present and thus the reconstruction is made with higher order interpolation, has larger density than the surface, and hence, the density-weighted result should show a higher convergence order.

\begin{figure}
	\begin{center}
		\includegraphics[height=210mm]{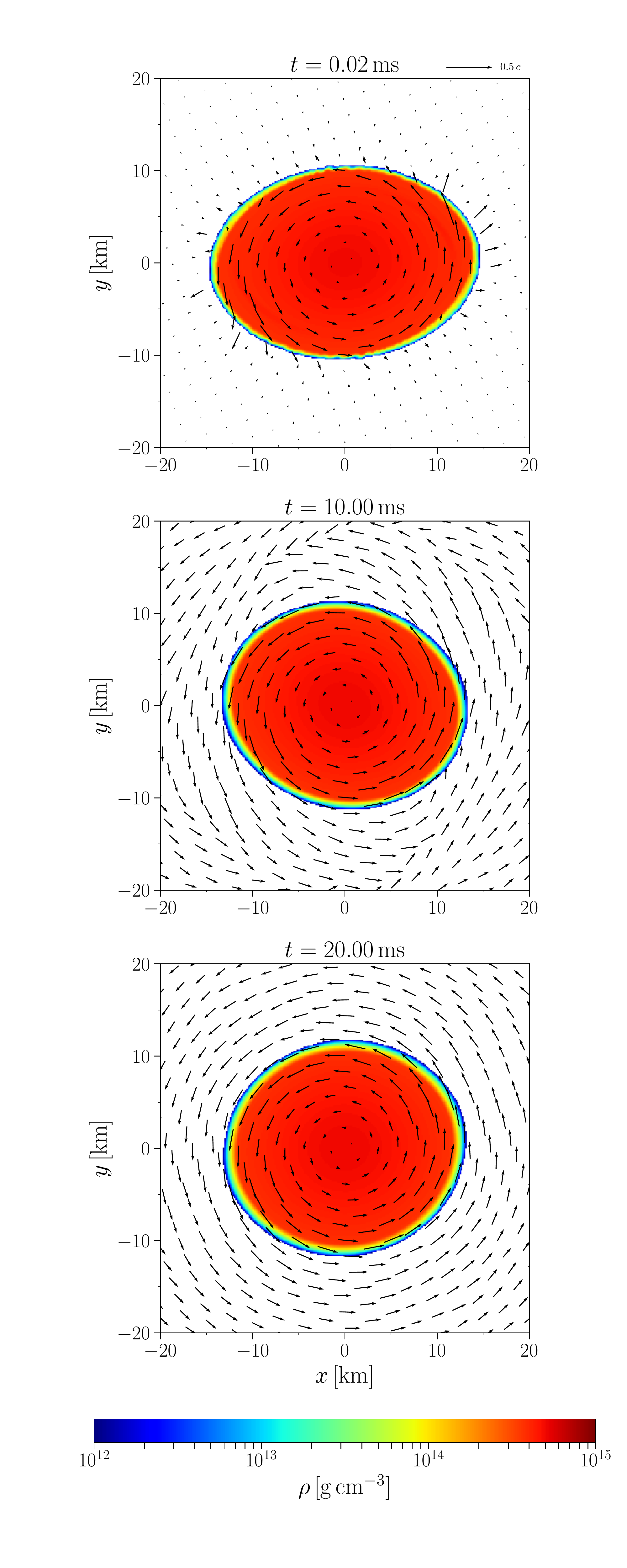}
	\end{center}
	\caption{Contour plots of the rest-mass density $\rho$ on the $x$-$y$ plane for model MIT148 at $0.02$, $10.00$, and $20.00$\,ms after the beginning of the simulation. The rapid color change in the vicinity of the surface indicates the existence of the surface discontinuity. The arrows show the 3-velocity field of the fluid. The lower limit for the density in the colorbar is set to be $10^{12}\,\mathrm{g\,cm^{-3}}$ which is already two orders of magnitude lower than $\rho_\mathrm{s}$. For most of the part outside the star, density is as small as $10^{-12}\rho_\mathrm{s}$ as explained in Sec.~\ref{sec:nm}.}
	\label{fig:snapshot148}	
\end{figure}

\begin{figure}
    \begin{center}
    	\includegraphics[height=210mm]{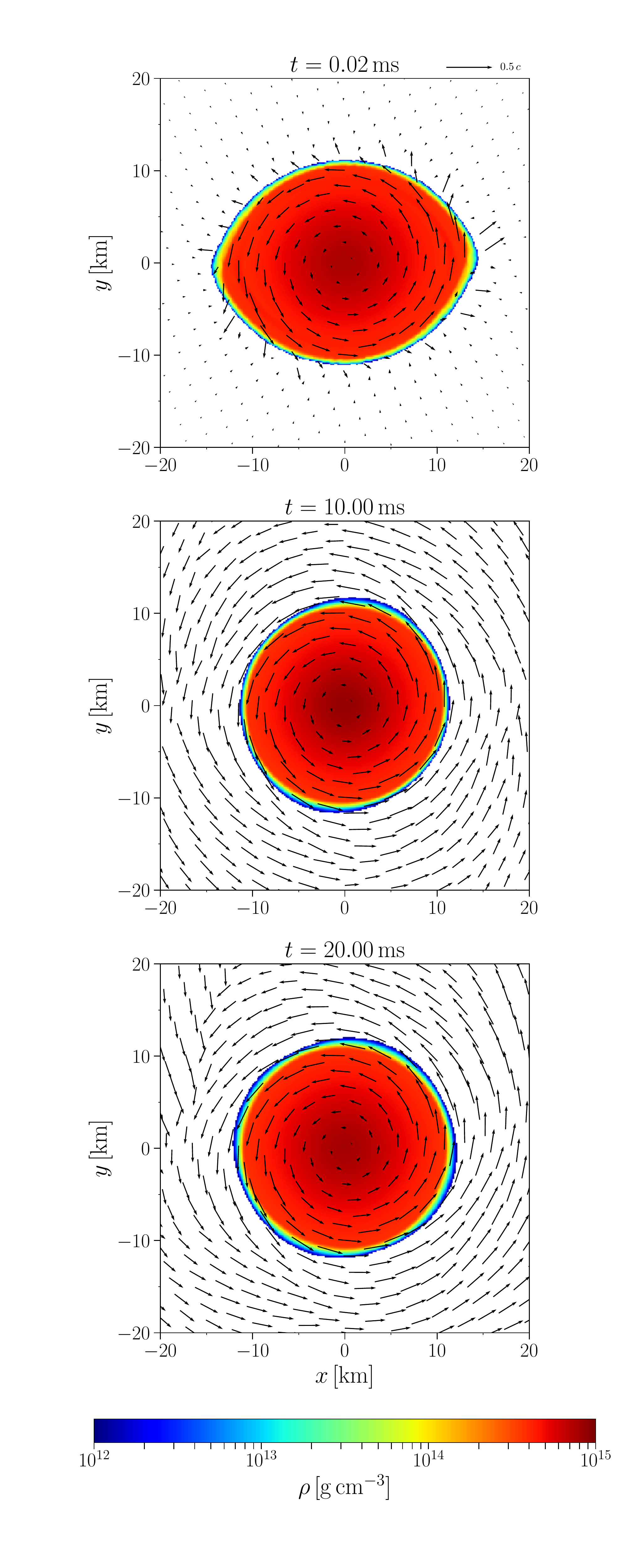}
	\end{center}
	\caption{Same as Fig.~\ref{fig:snapshot148} but for model MIT265.}
	\label{fig:snapshot265}	
\end{figure}

Apart from the relatively-low convergence order, we emphasize that our method for resolving the discontinuous density across the surface of the QS shows no sign of the problem. To clarify this point, we generate contour plots for the rest-mass density $\rho$ on the $x$-$y$ plane together with the 3-velocity field (Figs.~\ref{fig:snapshot148} and~\ref{fig:snapshot265}). Figures illustrate that the primitive recovery scheme we introduced works well for QSs. First it is found that the density drops from the surface density of the QS (which is approximately $3.737\times10^{14}\,\mathrm{g\,cm^{-3}}$ ) for two orders of magnitude within $\sim 3$ grid points; this is indicated by the rapid color change from orange to green, blue, and eventually white color within 3 grid points near the surface. We note that this sharp discontinuity is still resolved by $\approx 3$ grid points after 20\,ms as in the beginning of the simulations for both models (see the lower panels of Figs.~\ref{fig:snapshot148} and~\ref{fig:snapshot265}). No sign of the diffusion of this finite surface density is observed. Secondly, the velocity field, which is related to the Lorentz factor $w$ obtained by the primitive recovery, shows a smooth configuration as found from the figures; the velocity field inside the QSs follows a rigid bulk rotation and the outside does not (since we did not put any initial velocity for the atmosphere, as found from the upper panels of Figs.~\ref{fig:snapshot148} and~\ref{fig:snapshot265}). Especially, if we focus on the points near the surface of the QSs, it is found that the velocity on any points that are inside the surface follows the bulk motion of the QS while on the points one grid outside deviate. Since the surface of the QSs is moving in the domain (i.e., points inside/outside the QSs at one time step might move to their outside/inside at the next time step), the fact that the velocity field is well recovered for the main body of the QSs indicates the success of our implementation.

\section{Results}
\label{sec:results}
\subsection{Triaxial rotation case}
It has been suggested previously that a fast rotating compact star formed during collapse of a massive star or merger of a binary system could actually be in a configuration similar to those of Jacobi ellipsoids \citep{Chandrasekhar:1970}. The evolution of such configurations as well as the GW emission properties have been investigated both analytically and numerically \citep{Chandrasekhar1969book,Chandrasekhar:1970,Ipser1984,Lai95,Tsokaros2017}. According to these studies, the general understanding is that due to angular momentum loss by GW radiation, such triaxial 
star settles towards an axisymmetric configuration. The frequency of GWs is twice the spin frequency. In this stage, the angular velocity of the star, i.e., the GW frequency, increases as a result of the decrease of the moment of inertia during the reconfiguration of the triaxial star. When the star approaches the axisymmetric bifurcation point, other secular instabilities such as Dedekind instability and Chandrasekhar-Friedman-Schutz instability could be induced \citep{Chandrasekhar:1970,Lai95}. However, such instabilities only take place in a timescale much longer than the dynamical one as
\beq
\tau_\mathrm{gw}\sim2\times10^{-5}\left(\frac{M}{1.4\,M_\odot}\right)^{-3}\left(\frac{R}{10\,\mathrm{km}}\right)^{4}\left(\beta-\beta_\mathrm{sec}\right)^{-5}\,\mathrm{s},
\eeq
where $\beta$ and $\beta_\mathrm{sec}$ are the $T/|W|$ ratio of the star and that of the onset of the secular instability, respectively. Although QS models indeed could reach higher $T/|W|$ ratio than NSs and so is the case of the two models considered in this paper, this timescale is still as long as $\sim$40\,s for MIT265 and $\sim$2000\,s for MIT148, which are much longer than the timescale that can be covered by NR simulations. Hence, the simulation for secularly unstable stars is beyond the topic of this paper. On the other hand, the dynamical timescale is approximately written as
\beq
\tau_\mathrm{dyn}=\left(\frac{GM}{R^3}\right)^{-1/2},
\eeq
which is the same order of magnitude as the spin period, $\alt 1$\,ms, of the models we considered and much shorter than the timescale that can be covered by NR simulations. Therefore, we can conclude that triaxial QSs are dynamically stable, if the QSs still maintain their structure after tens of rotation periods.

\begin{figure}
	\begin{center}
		\includegraphics[height=70mm]{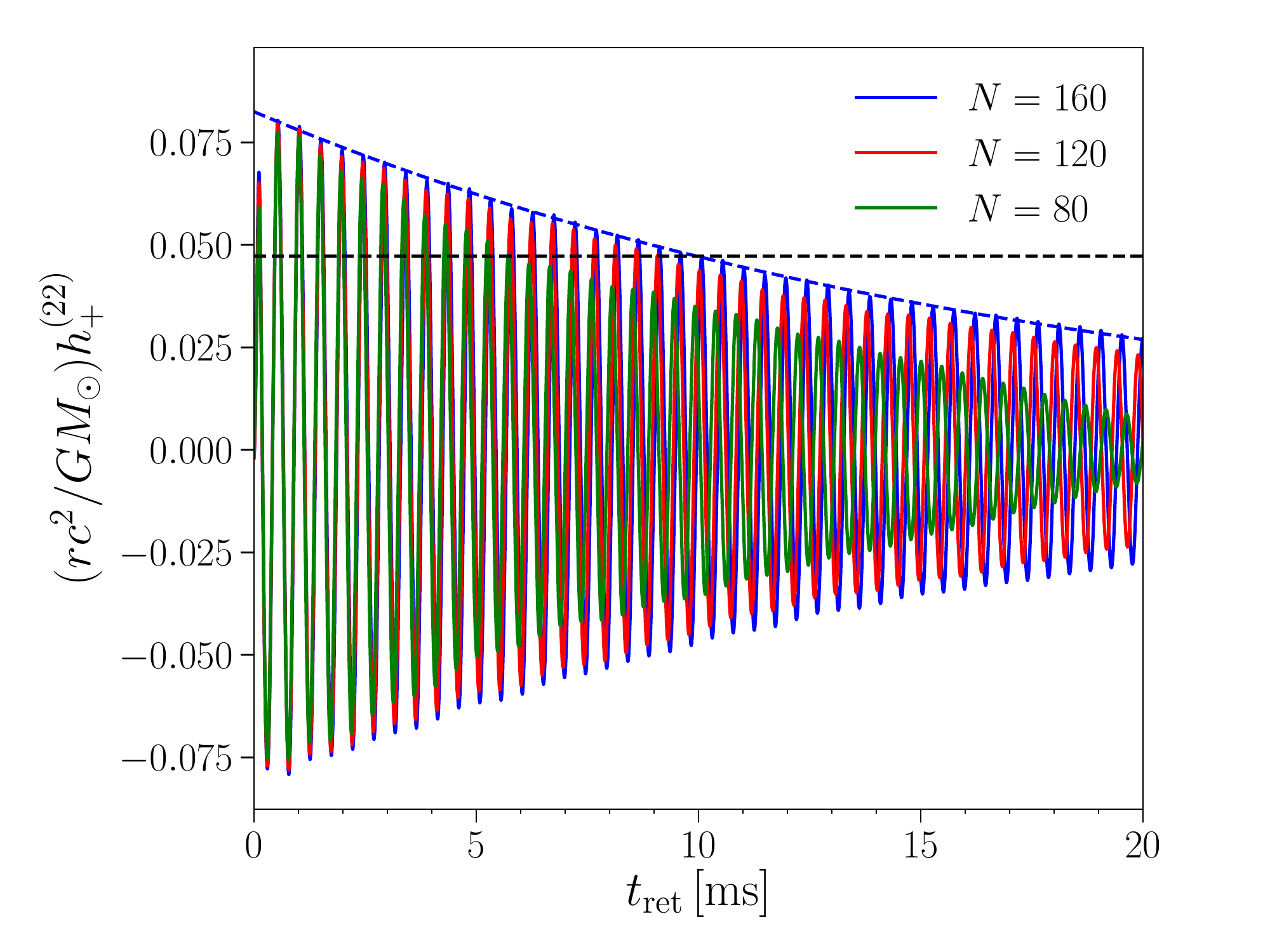}
		\includegraphics[height=70mm]{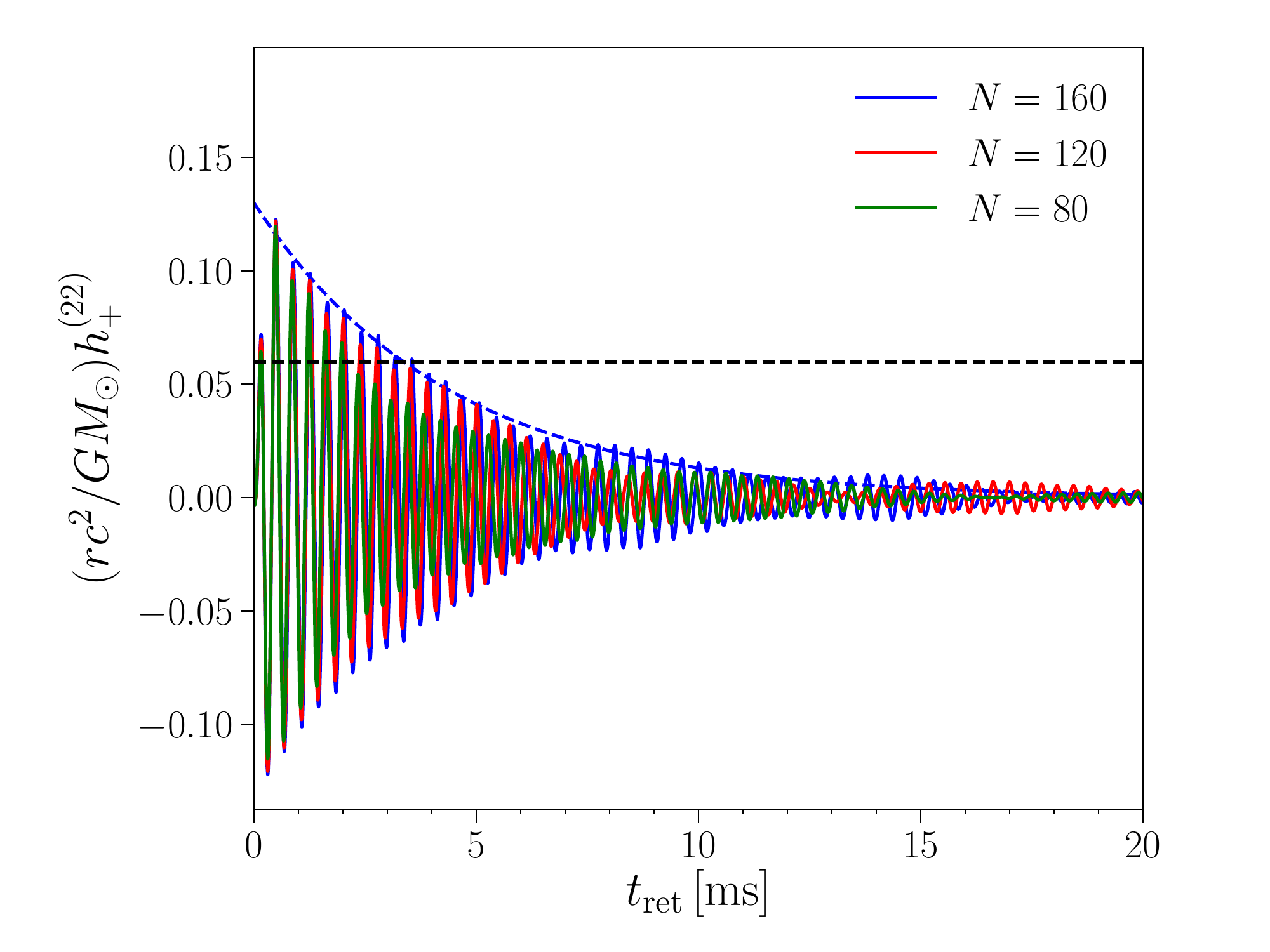}
	\end{center}
	\caption{Distance-normalized GW strain for the $h_{+}^{(22)}$ component extracted from the simulations for models MIT148 (upper panel) and MIT265 (lower panel). Results from three different resolutions are shown by the solid lines with different colors (blue for $N=160$, red for $N=120$, and green for $N=80$ case). The black-dashed horizontal line indicates the GW strain estimated by the quadrupole formula according to the initial data to show consistency. We find that the norm of the GW strain (i.e., $\sqrt{h_{+}^2+h_{\times}^2}$) can be well fitted with an exponential decay. We show the fit according to the best resolution results with the blue dashed curve. The best fitting exponential decay timescale is 17.85\,ms for MIT148 and 4.35\,ms for MIT265. The fixed frequency integration method \citep{Reisswig:2011} is used for obtaining the GW strain. The time shown here is the retarded time $t_\mathrm{ret} \sim t-1.5\,$ms as GW signal is extracted at $r=300\,GM_\odot/c^2$.}
	\label{fig:gw}
\end{figure}

Figures~\ref{fig:snapshot148} and~\ref{fig:snapshot265} indicate that the configuration of the QSs settles towards an axisymmetric one. Quantitatively speaking, after 10\,ms, the length of the equatorial radius along the longer axis shrinks to 12.4\,km (which is $11\%$ shorter than the initial value of $R_x$) for model MIT148. On the other hand, for MIT265, $R_x$ shrinks to 10.6\,km and becomes approximately identical to $R_y$. This implies that the QSs lost nearly all the triaxial deformation in the first 10\,ms. As the triaxial deformation decays during the evolution, the same occurs in the GW amplitude. As shown in Fig.~\ref{fig:gw}, specifically, the GW strain amplitude decays exponentially with time. This indicates that the GW dissipation would be indeed the dominant mechanism for the reconfiguration of the triaxial QSs. 

\begin{figure}
	\begin{center}
		\includegraphics[height=70mm]{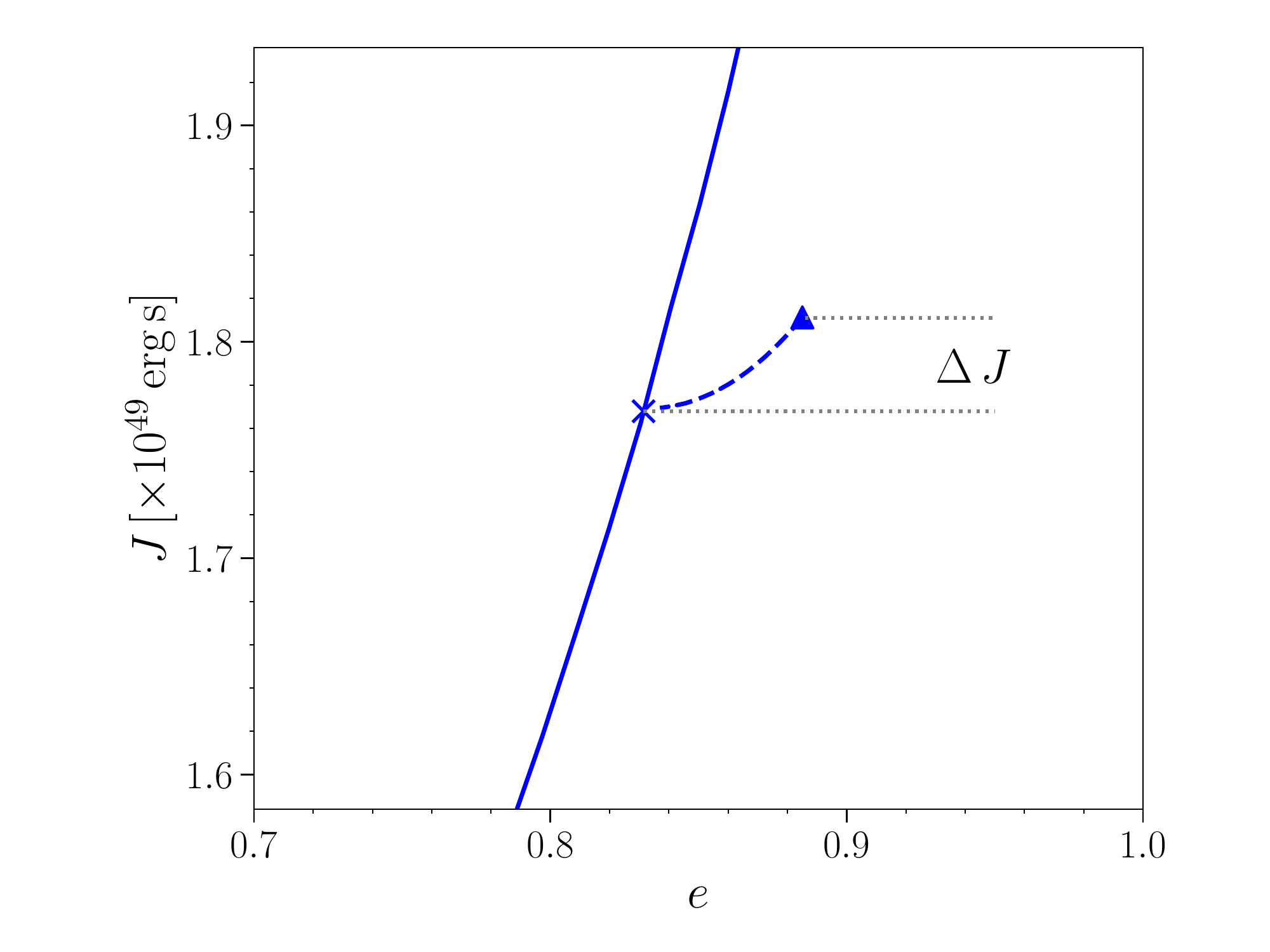}
	\end{center}
	\caption{Illustration of the difference between $\Delta J$ and total angular momentum for triaxially rotating solutions. The eccentricity ($e=\sqrt{1-(R_z/R_x)^2}$, where $R_z$ and $R_x$ are the coordinate lengths of the polar and equatorial radii) and angular momentum $J$ of a solution is shown for both axisymmetric solutions (solid line) and triaxial solutions (dashed line) with the same baryonic mass as model MIT148 considered in this work. The triangle and cross indicate model MIT148 and the bifurcation point, respectively. There exists no triaxial solutions if the angular momentum is smaller than that of the bifurcation point. Therefore, after losing the extra angular momentum with respect to the bifurcation point ($\Delta J$ in the figure), axisymmetry is resumed for MIT148 model.}
	\label{fig:deltaj}
\end{figure}

A similar numerical result was found previously for NSs~\citep{Tsokaros2017}. The authors in \citep{Tsokaros2017} suggested that the timescale for reconfiguration of the stars might be incompatible with the estimation of the GW dissipation timescale that they assume:
\beq
\tau_\mathrm{gw,dis}=T/\dot{E},
\label{eq:gwdis}
\eeq
where $T$ is the rotational kinetic energy of the star and $\dot{E}$ is the GW luminosity. They showed that this timescale is of the order of 1\,s, and hence, it cannot account for the change in the stellar shape that proceeds in a timescale of 10\,ms. However, Eq.~\eqref{eq:gwdis} is not appropriate for estimating the relevant timescale and actually overestimates the timescale by a factor of $\sim 100$. The reason for this is that 
the triaxial stars only need to lose the extra angular momentum (or extra rotational kinetic energy) relative to that of the bifurcation configuration, rather than all their angular momentum (or rotational kinetic energy), to settle to the axisymmetric configuration. Therefore, a better estimation should be given by
\beq
\tau_\mathrm{gw,dis}=\Delta J/\dot{J},
\eeq
where 
$\Delta J$ is the angular momentum difference between the initial configuration and that of the bifurcation point solution (see the illustration in Fig.~\ref{fig:deltaj}) and $\dot{J}$ is the angular momentum loss rate due to the GW emission which is related to $\dot{E}$ approximately as $\dot{J}\approx \dot{E}/\Omega$. According to the previous studies~\citep{Zhou2018,Huang08}, triaxial stars can only obtain maximumly $10\%$ extra angular momentum relative to that of the bifurcation point, $J_b$. In particular, for the stars with the larger compactness, the smaller amount of the extra angular momentum can be reached (cf.~Fig.8 in~\citep{Zhou2018} or Table~V in~\citep{Huang08}). For the high compactness case, the extra angular momentum could be as small as $\sim1\%$ of $J_b$. Therefore, it is reasonable that the triaxial stars settle to an axisymmetric configuration in a timescale of tens of milliseconds by the GW emission. In fact, according to the exponential fit shown in Fig.~\ref{fig:gw}, this timescale is 17.85\,ms for MIT148 and 4.35\,ms for MIT265. These values are consistent with the argument that $\Delta J$ is smaller for higher-compactness QSs.

\begin{figure}
	\begin{center}
		\includegraphics[height=70mm]{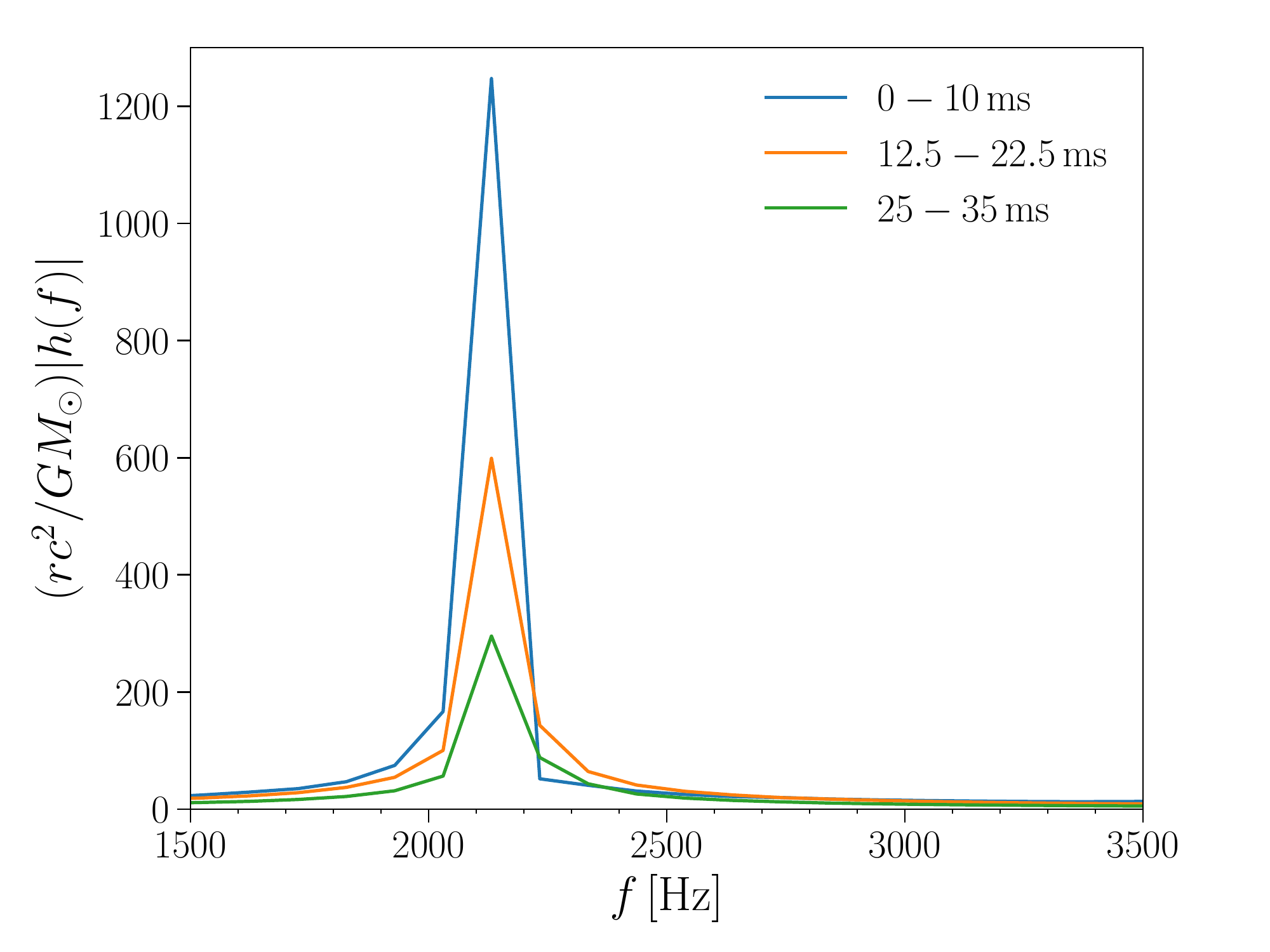}
		\includegraphics[height=70mm]{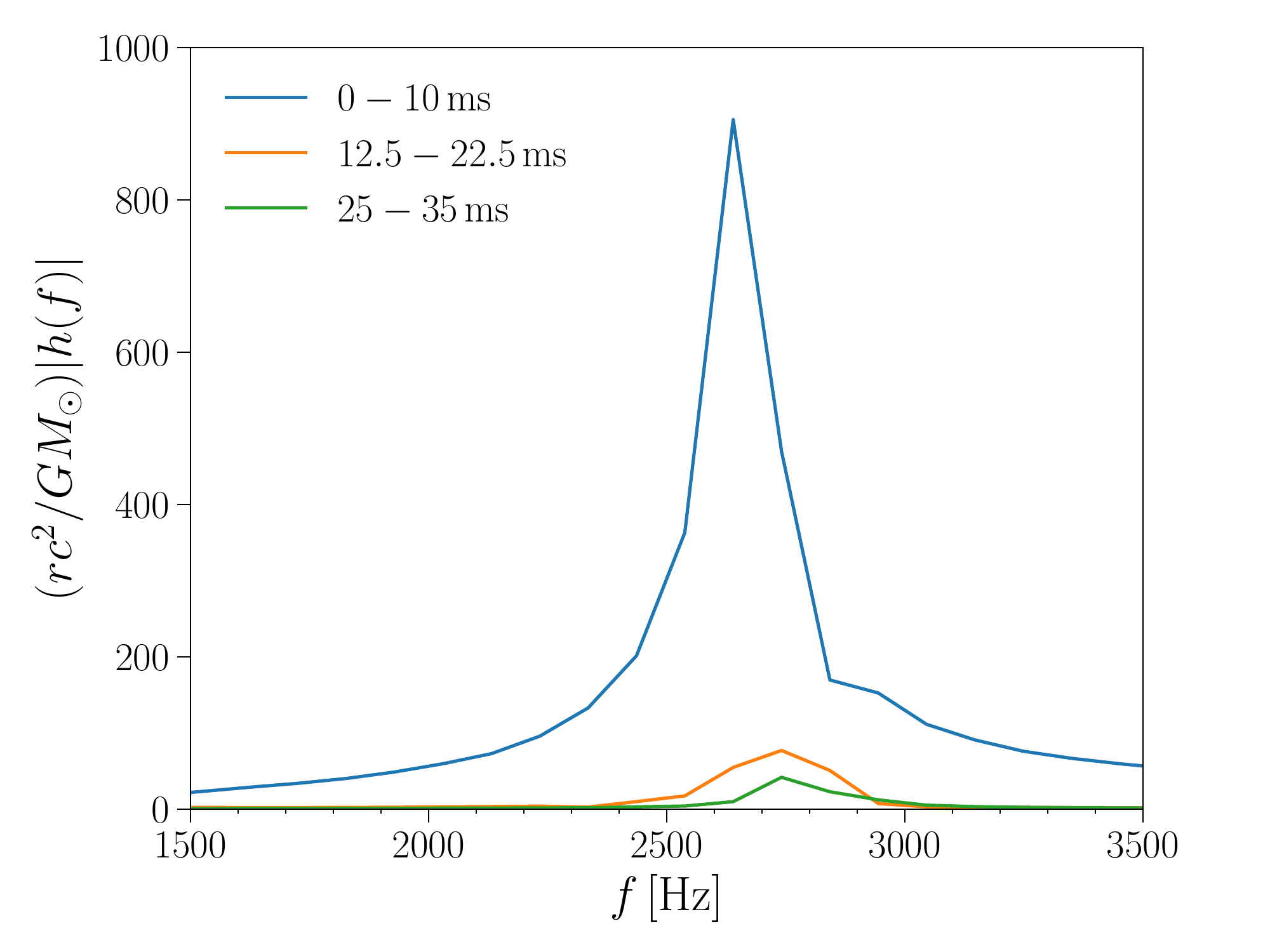}
	\end{center}
	\caption{Distance-normalized power spectrum of GW for models MIT148 (upper panel) and MIT265 (lower panel). The Fourier transformation is performed at different time bins to study the evolution of the dominant frequency. The time indicated is the retarded time as in Fig.~\ref{fig:gw}.}
	\label{fig:gwf}
\end{figure}

The decay of the triaxial deformation results not only in the damping of the GW amplitude but also in the shift of the GW frequency, as the triaxial star spins up a little bit. We analyze the evolution of the GW frequency by performing Fourier transformation for the GW strain in different time intervals. The result is shown is Fig.~\ref{fig:gwf}. For model MIT148, the GW signal is approximately monochromatic with the dominant frequency at $f=2.14\,$kHz \footnote{Note that the time window has a size of 10\,ms for performing the Fourier transformation, the resulting frequency resolution is 100\,Hz, and thus, in principle, we cannot resolve if the frequency shift is smaller than 100\,Hz.} up until 35\,ms, although the power decreases by a factor of 5 in $\sim 30$\,ms. On the other hand, for model MIT265, the frequency of the dominant peak shifts from $f=2.64\,$kHz in the first time bin to $f=2.74\,$kHz in the second time bin. 
The comparison between models MIT148 and MIT265 indicates that due to angular momentum dissipation by GW radiation, the reconfiguration of the triaxial star toward axisymmetry is faster for models with higher mass, which is
again a result of the fact that higher compactness stars contain much less extra angular momentum relative to that of the axisymmetric bifurcation point. 

In addition, it is worth noting that, although the instantaneous GW amplitude in the beginning for model MIT265 is larger than that for model MIT148 (same occurs for the GW amplitude estimated in the initial data), it is unlikely that supramassive triaxial QSs are stronger GW sources. This is again due to the fact that the decay timescale of the GW amplitude is too short for the supramassive cases, and in the realistic detection, the signal to noise ratio could be less gained. This fact can be directly understood from Fig.~\ref{fig:gwf}, as the power spectrum is indeed stronger at any given time bin for model MIT148 than for MIT265. At any rate, the signal still decays too fast to become ideal continuous GW sources even for MIT148 unless the triaxial configuration could be maintained by external angular momentum supplies (by accretion for example), known as the forced GW emission scenario \citep{Wagoner1984,Papaloizou1978}.

\subsection{Differential rotation case}
\label{sec:drot}

It is analytically shown that an incompressible Maclaurin spheroid becomes dynamically unstable to bar formation in Newtonian gravity if $T/|W|$ is larger than a critical value of $\sim0.27$ \citep{Chandrasekhar1969book}. Such a high $T/|W|$ ratio could be reached by rapidly spinning NSs and QSs only in the presence of differential rotation. Previous studies have shown that also in GR, the dynamical bar-mode instability could be induced for differentially rotating NSs with a sufficiently high value of $T/|W|$, and the critical value is found to be slightly smaller than the Newtonian value ($\sim 0.25$ in GR) depending weakly on the stiffness of the EoS models \citep{Shibata:2000jt,Baiotti06b,Loeffler2015,Pietri2014}. It has been shown that for such NSs, bar-like perturbation grows exponentially in the early stage until saturation is reached. Beyond the saturation, the evolution differs for models with different $T/|W|$ values: spiral arms are formed and mass ejection subsequently occurs for relatively large values of $T/|W|$, while for models with $T/|W|$ close to the critical value, no significant sign of the spiral arm structure and mass ejection is found \citep{Shibata:2000jt}. In this paper, both MIT275dr and APR206dr possess $T/|W|$ larger than 0.26. Thus, we expect that the spiral arm structure is formed. These models could be also useful for understanding the similarity or difference between the bar-mode instability of QSs and NSs.

\begin{figure}
	\begin{center}
		\includegraphics[height=70mm]{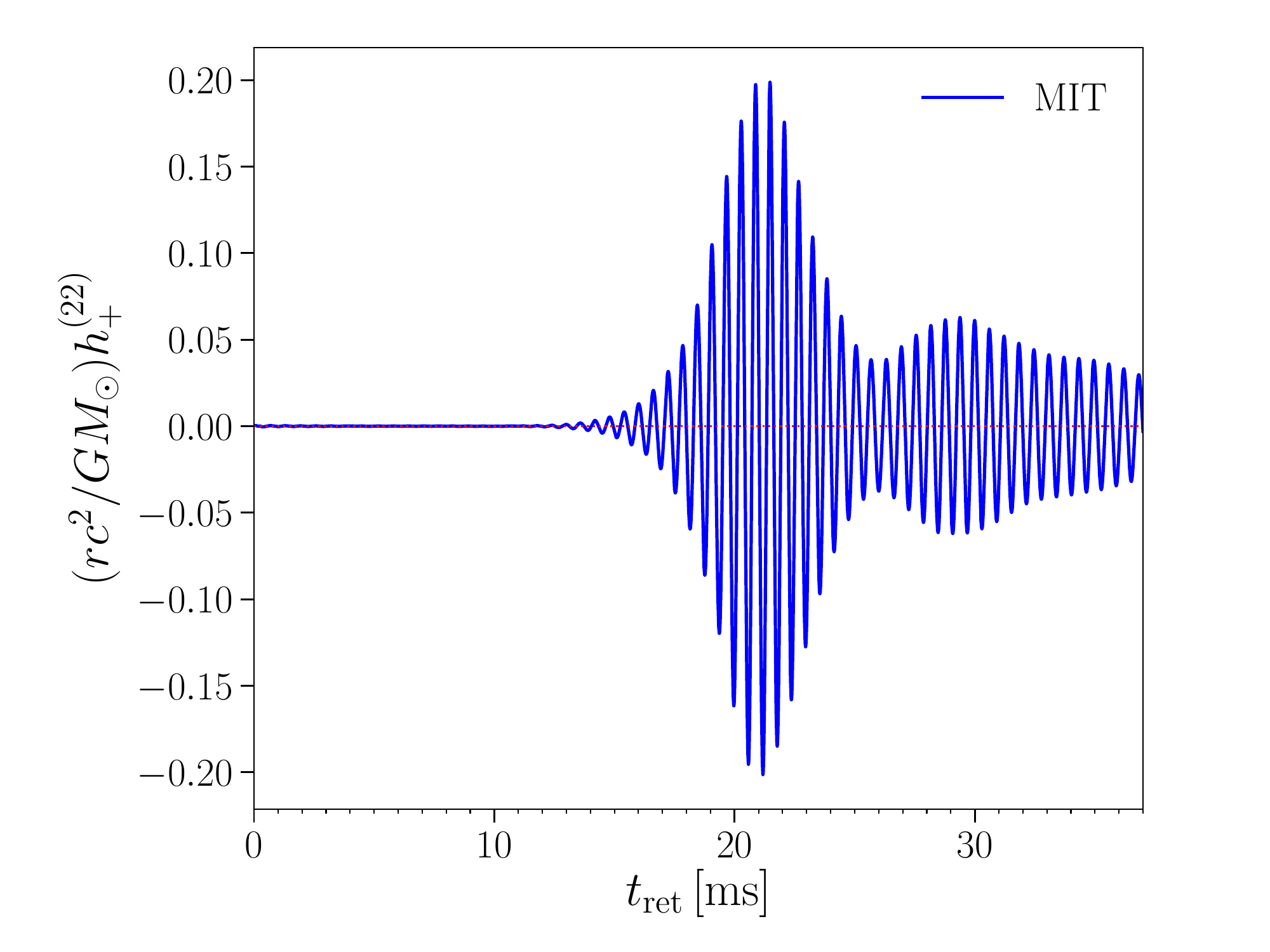}
		\includegraphics[height=70mm]{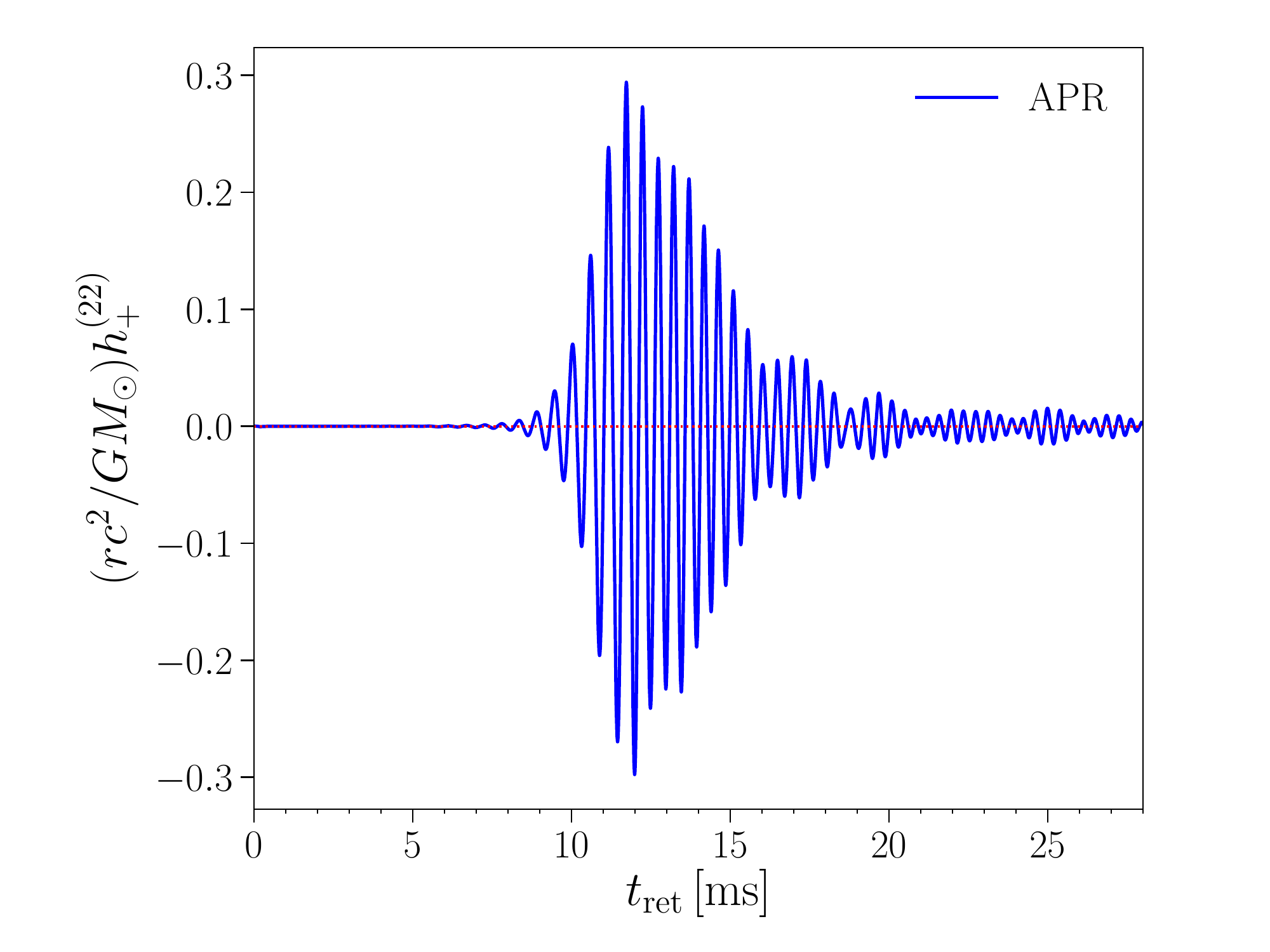}
	\end{center}
	\caption{Distance-normalized GW strain for the $h_{+}^{(22)}$ component extracted from the simulations for models MIT275dr (upper panel) and APR206dr (lower panel) in the highest resolution case.}
	\label{fig:gwdr}
\end{figure}

The $l=m=2$ mode GW strain, which is directly related to $m=2$ bar-mode, during the evolution is shown in Fig.~\ref{fig:gwdr}. It is worth noting that we \emph{do not impose any bar-like perturbation} in the beginning of the evolution, which is different from the treatments in previous studies \citep{Shibata:2000jt,Baiotti06b} as our main focus is not to explore the parameter space for un-/stable differentially rotating QSs, but to understand the capability of the code as well as the difference between QSs and NSs. In our evolution, bar-mode instability \emph{spontaneously} sets in from a random perturbation that should present in any numerical simulation, and exponentially grows from the tiny perturbation spending $\sim10$--$20\,$ms for both models. Saturation is achieved approximately at $t_\mathrm{ret}=19.75\,$ms for model MIT275dr and at $t_\mathrm{ret}=12.25\,$ms for model APR206dr, and then, the GW strain decreases to a relatively lower level.

\begin{figure}
	\begin{center}
		\includegraphics[height=210mm]{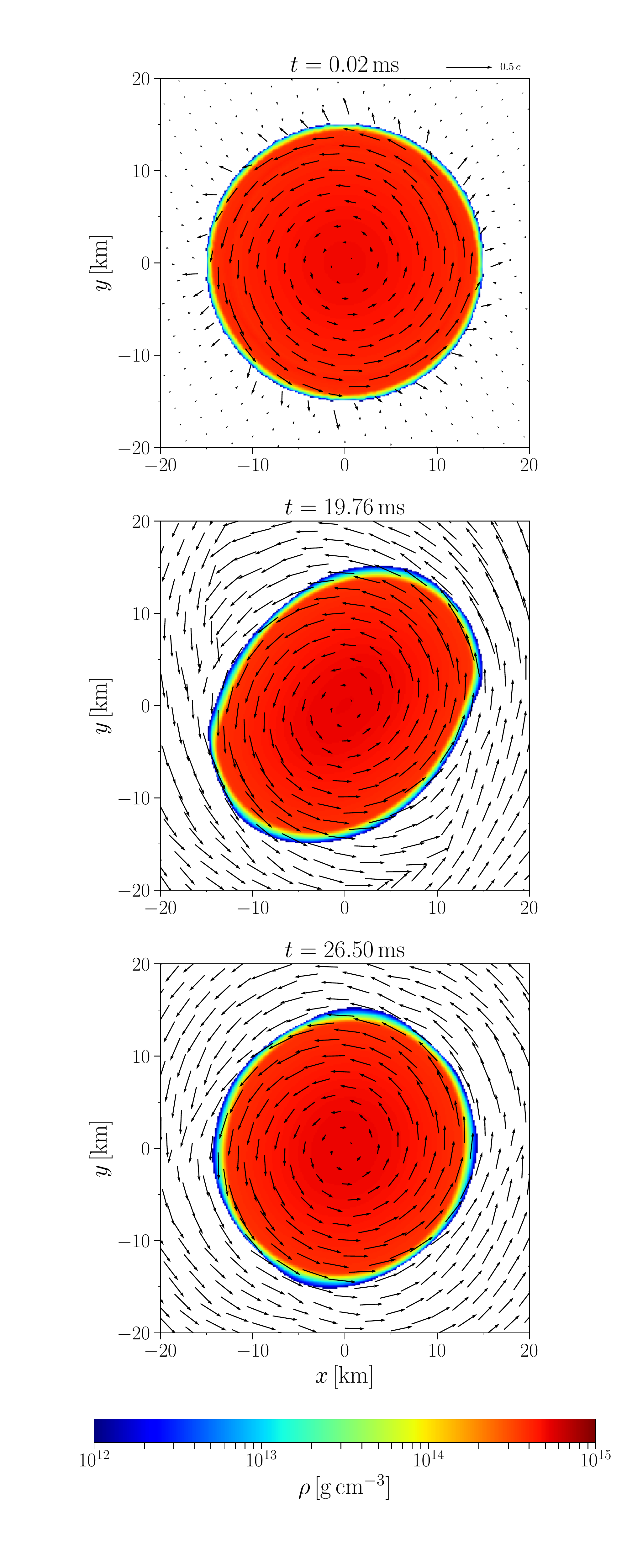}
	\end{center}
	\caption{Contour plots of the rest-mass density $\rho$ on the $x$-$y$ plane for model MIT275dr at $0.02$, $19.76$, and $26.50$\,ms after the beginning of the simulation. Note that at $t_\mathrm{ret}=19.76\,$ms, the maximum value of $h_{+}^{(22)}$ is achieved during the simulation (cf. Fig.~\ref{fig:gwdr}) and at $t_\mathrm{ret}=26.50\,$ms, the GW amplitude decreases to a relatively low level.}
	\label{fig:snapshot275}	
\end{figure}

\begin{figure}
	\begin{center}
		\includegraphics[height=210mm]{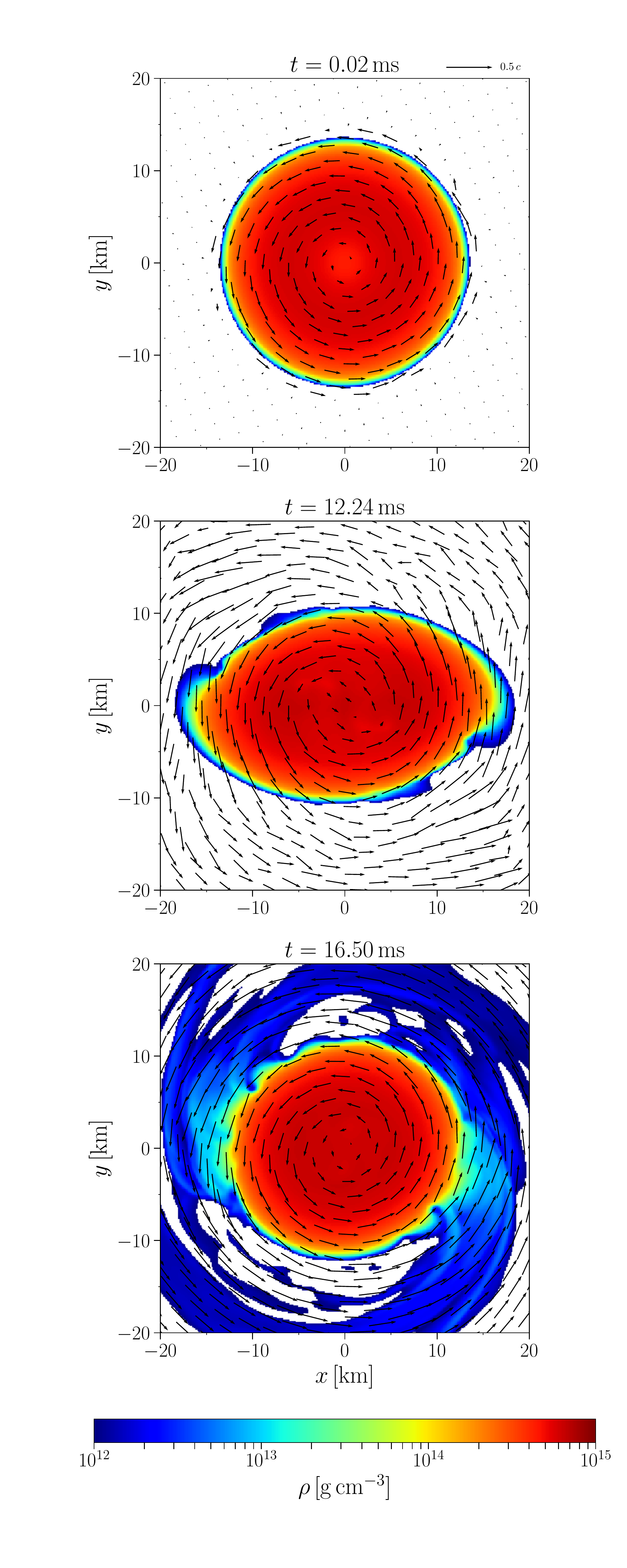}
	\end{center}
	\caption{Same as Fig.~\ref{fig:snapshot275} but for model APR206dr. The time of the snapshots is selected in the same way as that in the caption of Fig.~\ref{fig:snapshot275}. }
	\label{fig:snapshot206}	
\end{figure}

The snapshot of density contours on the equatorial plane straightforwardly demonstrates the growth of the bar mode (see Figs.~\ref{fig:snapshot275} and \ref{fig:snapshot206}). The evolution of the density contours for model APR206dr is quite consistent with the previous results of differentially rotating compact stars with a relatively large value of $T/|W|$ \citep{Shibata:2000jt,Baiotti06b}: the star is significantly distorted when the bar-mode growth saturates (the middle panel, which we choose the time when the peak of the GW strain is reached) and the spiral arm structure forms afterwards (the lower panel). On the contrary, the result of model MIT275dr is similar to the previous results with $T/|W|$ slightly larger than the critical value: the star adjusts to ellipsoidal structure without formation of spiral arms. 

\begin{figure}
	\begin{center}
		\includegraphics[height=70mm]{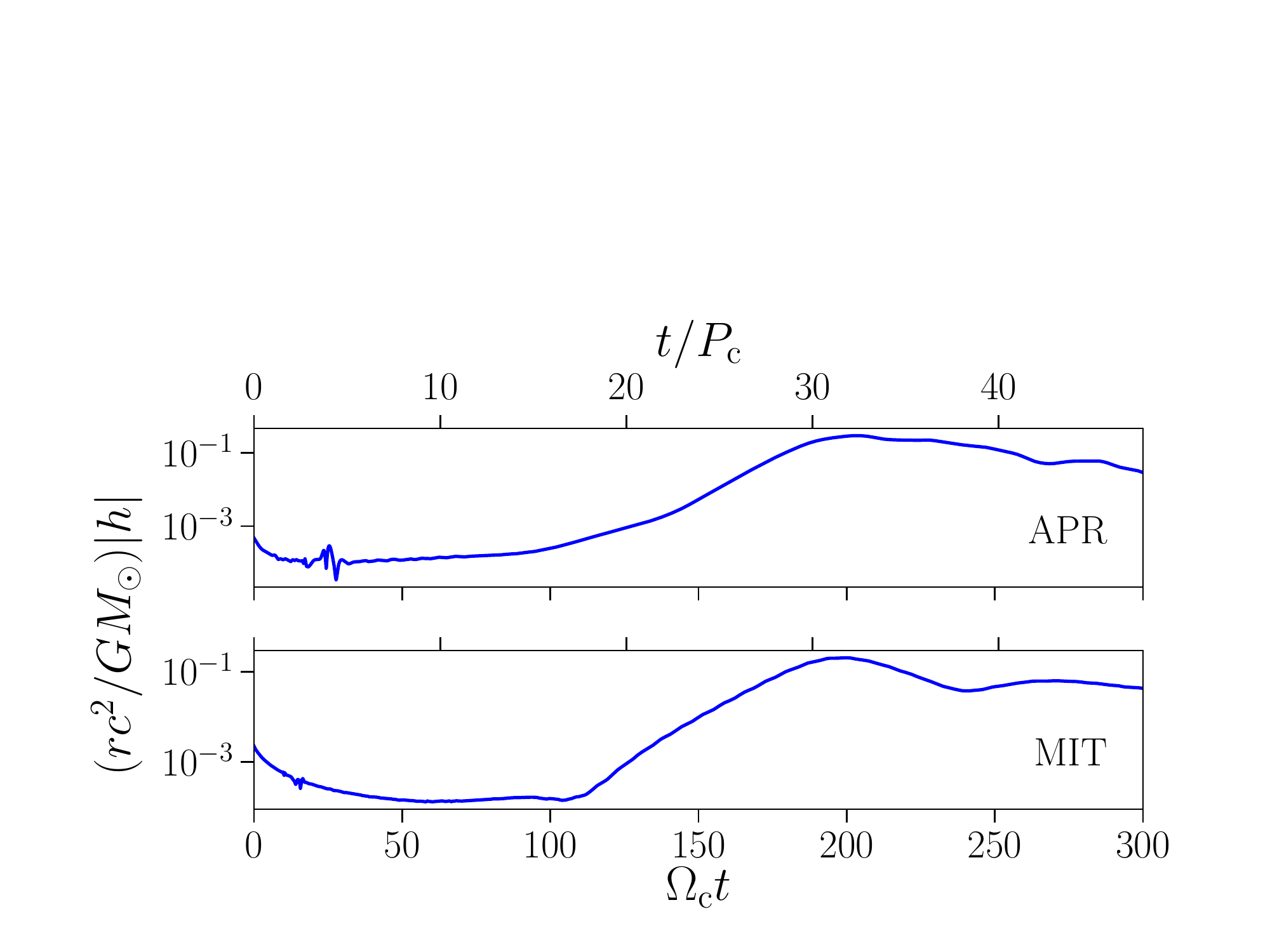}
	\end{center}
	\caption{Distance-normalized GW strain norm ($|h|=\sqrt{h_{+}^2+h_{\times}^2}$) for the $l=m=2$ component extracted from the simulations for models MIT275dr (lower panel) and APR206dr (upper panel) in the highest resolution case. We have multiplied the time by the central angular velocity (shown as the $x$-axis in the bottom) of the two models, respectively, such that the horizontal axis is now proportional to the number of central rotation periods (shown as the $x$-axis on the top).}
	\label{fig:gwnormdr}
\end{figure}

In spite of all the differences mentioned above, it is found that the growth rate of the bar-mode instability is quite similar for the two models considered in this work if we normalize the simulation time by the central rotation frequency (cf. Fig.~\ref{fig:gwnormdr}). For both models, the GW amplitude grows for about 3 orders of magnitude in the duration of $\Omega_\mathrm{c}t=100$--$200$. More careful analysis yields an $e$-folding timescale of about 2.24 central rotation periods for model APR206dr and 2.04 for model MIT275dr. This result is also consistent with the range derived in previous studies~\citep{Shibata:2000jt,Baiotti06b}, indicating that the instabilities found in this work are indeed the dynamical bar-mode one. 

\begin{figure}
	\begin{center}
		\includegraphics[height=70mm]{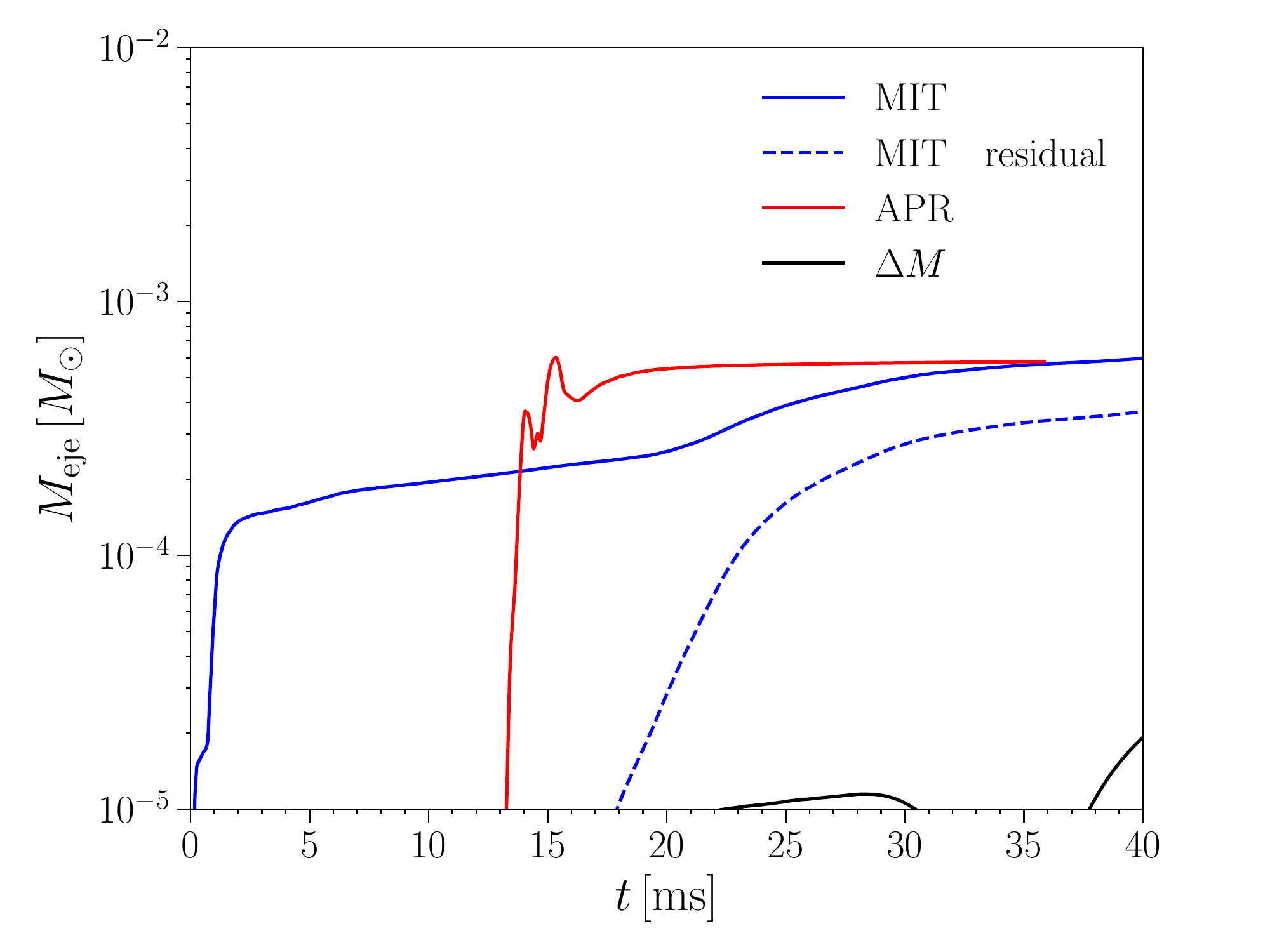}
	\end{center}
	\caption{Amount of mass ejected according to the criterion $u_t<-1.0$ for models APR206dr (red solid line) and MIT275dr (blue solid line) during the evolution. To remove the effect of artificial mass ejection in the early stage of the QS case, we have subtracted out the mass ejection before $t=18\,$ms and the residual is shown as the blue dashed curve. The black curve indicates the absolute value of change in baryonic mass for model MIT275dr. When matter is ejected and moving out of the finest FMR level to coarser levels, baryonic mass conservation might be violated but still stays  by one order of magnitude lower than the amount of ejected material.}
	\label{fig:ejedr}
\end{figure}

The mass ejection is also studied (as shown in Fig.~\ref{fig:ejedr}) by applying the criterion that matter becomes unbound if $u_t$ is smaller than $-1.0$. For model APR206dr, a burst of mass ejection is found approximately at the same time when the GW strain reaches the maximum value and in total about $6\times10^{-4}\,M_\odot$ is ejected in the end. However, the situation is a bit more complicated for the QS case due to the artificial mass ejection in the beginning. The main reason for this artificial mass ejection is the existence of matter with density smaller than the surface density near the surface when initial data is interpolated to the grid points for dynamical evolution. Such matter has to be treated as ideal gas once the dynamical evolution started and this may drive part of those matter away from equilibrium. Nevertheless, a rise in the ejected mass is still found at $t\sim20\,$ms, which corresponds to the duration of the growth in the GW strain. A total mass ejection of about $3.5\times10^{-4}\,M_\odot$ is found after subtracting the mass ejection before $t=18\,$ms (i.e., before the bar-mode begins to exponentially grow). A more careful study on this artificial ejection is shown in Appendix~\ref{sec:appeje}. It is found that, for the current implementation, the quantitative measurement of the ejecta is reliable for slowly rotating models but uncertainty will significantly increase for models with extremely large angular velocity (as is the case for MIT275dr) due to artificial ejection.

An interesting feature is that although model MIT275dr possesses a higher $T/|W|$ value than model APR206dr (cf.~Table~\ref{tab:id}), the spiral arm formation and mass ejection are much less significant than the NS case, showing that the critical $T/|W|$ value is likely to be higher for the QS model we considered here. There are several possible reasons for this. First of all, for QSs the density profile is much more uniform than for NSs, and hence, the $T/|W|$ ratio is possible to be higher to approach the limit for incompressible stars. Secondly, previous studies concentrate in a differential rotation degree around $\hat{A}=1.0$ for NSs and found weak correlation between critical $T/|W|$ and $\hat{A}$ parameter. However, for MIT275dr model considered here, a quite large $\hat{A}$ parameter (3.0) is adopted and for NSs with such small differential rotation degree it is not possible to find dynamically unstable solutions. The parameter space for dynamical bar-mode instability for such large $\hat{A}$ parameter might be different and we need to perform more systematic studies for differentially rotating QSs with different parameters to fully understand this in the future. 

\section{Discussion and Conclusion}
\label{sec:disandconclu}
In this paper, we introduced our approach to perform NR simulations for QSs which have a finite surface density.  We made two major efforts for successfully implementing QSs described by the MIT bag model into our NR code \SACRA. First, we appropriately chose the EoS parameters in such a way to make specific enthalpy continuous and monotonic across the surface of the QS, without changing any quantities of the QSs described by this model. Secondly, we developed a procedure for the primitive recovery suitable for QSs. Specifically, in our approach, the primitive recovery is performed separately for matter inside and outside QSs, and in addition, we introduced an analytical approach for the QS part. We then examined the implementation by performing simulations for two triaxial QS models with different mass and a differential rotation model which is dynamically unstable to the bar-mode deformation. The surface of the QSs with the finite density is found to be well resolved for the entire time during the evolution and the primitive recovery is correctly executed for matter both inside and outside the QSs. Convergence studies revealed a first-order convergence as in the case of initial data construction for QSs~\citep{Zhou2018}. The success in capturing the evolving surface of QSs and mass ejection from it is also the important gain for us to move to the simulations for BQSs. Besides, this method can in principle be extended to perform simulations on compact stars with a first-order strong-interaction phase transition (for which there is a density jump inside the star). We plan to explore this topic in the future.

For the triaxially rotating QSs, the results are found to be qualitatively similar to the
case of NSs~\citep{Tsokaros2017}. The triaxial deformation of QSs decays as it radiates angular momentum by the GW emission in a timescale of $10\,$ms, which is found to be consistent with the GW dissipation timescale. As a result, the amplitude of GWs decays exponentially. It is also found that for higher compactness, the QS settles to an axisymmetric configuration faster, due to smaller amount of the extra angular momentum contained in comparison with that of the bifurcation point solution. We found that the GW emission of the model with typical mass of an individual compact star is approximately monochromatic, while the GW emission of the supramassive case shifts to a higher frequency within $\sim 10\,$ms as the star resumes axisymmetry at this time and the GW amplitude is significantly reduced. As a conclusion, in spite of the fact that rotating QSs could reach a $T/|W|$ ratio higher than that of NSs, and thus, supramassive triaxially rotating QSs do exist, it is unlikely that such supramassive triaxial rotating QSs could be a better source for continuous GWs, as the signal from such objects does not last for more than $\sim 10$\,ms.

For the case of differentially rotating stars, we confirmed that QSs with much smaller differential rotation degree could still experience the bar-mode instability as long as the $T/|W|$ ratios is large enough (i.e., $\agt 0.27$), as expected in our previous initial data studies~\citep{Zhou2019}. In particular, even without initially artificial bar-like perturbation, the bar-mode deformation of QSs still exponentially grows in a comparable timescale of the NS cases. The result is consistent with the previous results of NSs with $T/|W|$ ratios not much larger than the critical value~\citep{Shibata:2000jt,Baiotti06b}. This might be a result of the difference between the structure of QS and NSs or the fact that the $\hat{A}$ parameter considered here for QSs is quite different from previous studies of NSs. To fully understand the parameter space of bar-mode instability of differentially rotating QSs, a more systematic investigation needs to be done in the future.

\acknowledgments 
We thank Kyohei Kawaguchi, Elias Roland Most and the members of Computational Relativistic Astrophysics division in Max Planck Institute for Gravitational Physics (Potsdam) for very helpful discussions. This work is supported by the Ministry of Science and Technology of the People's Republic of China (2020SKA0120300), Grant-in-Aid for Scientific Research (Grant Nos. 16H02183, 18H01213, 20H00158 and 18K03624) of Japanese MEXT/JSPS. A. T. is supported by NSF Grants No. PHY-1662211 and No. PHY-2006066, and NASA Grant No. 80NSSC17K0070 to the University of Illinois at Urbana-Champaign. The simulations are performed on the COBRA and SAKURA at Max Planck Computing and Data Facility (MPCDF).

\appendix
\section{Tests of TOV Solutions}
\label{sec:apptov}
In order to more quantitatively understand the validity of our numerical implementation, it is helpful to investigate models which are simple but the evolution of them could be predictable. Evolution of TOV solutions, i.e., non-rotational solutions which should be static, is one good choice to 
understand the accuracy and convergence behavior of our methods quantitatively.

For this purpose, we have prepared evolution of a TOV solution with $M_\mathrm{ADM}=2.07\,M_\odot$, which is quite close to the TOV maximum mass of our EoS model, in 3 different resolutions. As introduced in Sec. \ref{sec:test}, nine FMR domains are used for the simulation and the half size of the finest level being  $8GM_\odot/c^2\approx11.8\,$km for this test run. There different resolutions with $\Delta_x=\Delta_y=\Delta_z=148,~98.4$, and $73.8\,$m are chosen, which corresponds to $N=80,~120$ and $160$ grid points along one direction.

\begin{figure}
	\begin{center}
		\includegraphics[height=70mm]{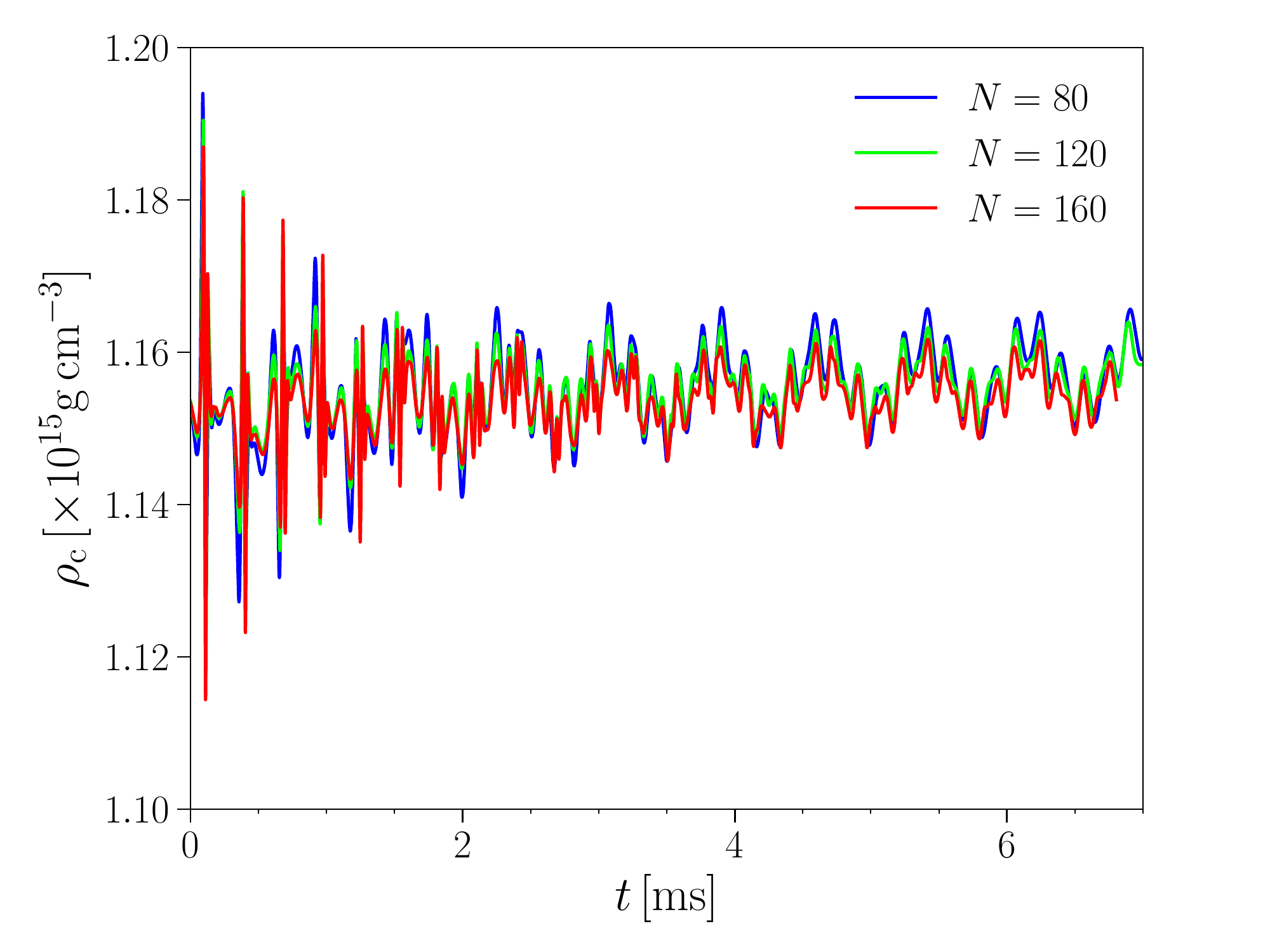}
	\end{center}
	\caption{Evolution of the central density for the test on the TOV solution case for 3 different resolutions.}
	\label{fig:rhoctov}
\end{figure}

Due to numerical perturbation which is present in any code, the central density will oscillate in dynamical timescale even for TOV solutions of NSs. Since we are considering the TOV solution whose mass is very close to its maximum, the QS will become unstable and collapse to BH if the numerical method is not reliable. Hence, it is important to check the convergence behavior on the oscillation of the central density for this test. As can be seen in Fig.~\ref{fig:rhoctov}, the oscillation amplitude of the central density decreases with time for all three resolutions. The amplitude is $\sim3\%$ in the beginning and then stabilize to roughly $1\%$ after 2\,ms. Particularly, the oscillation amplitude is smaller in higher resolution and can be used to estimate the convergence order. 

According to Eq.~(\ref{eq:convergence}), we can define
\begin{equation}
R=\frac{f_{160}-f_{120}}{f_{160}-f_{80}}=\frac{\Delta_{160}^{\zeta}-\Delta_{120}^{\zeta}}{\Delta_{160}^{\zeta}-\Delta_{80}^{\zeta}}=\frac{1-(4/3)^{\zeta}}{1-2^{\zeta}},
\end{equation}
in which $f_{160}, f_{120}$ and $f_{80}$ here represent the oscillation amplitude in three different resolutions, respectively. Since the oscillation phase could be different in 3 configurations and the amplitude could vary for every period, we have chosen roughly ten oscillation cycles from $t=3.5\,$ms to 5\,ms when the oscillation is already more stabilized for all 3 resolutions and define $f_N$ as
the difference between maximum and minimum value of the central density in this duration for the corresponding resolution. With such a definition, $R$ is found to be $0.2516$ which indicates that second order convergence is achieved for this quantity ($\zeta\sim2$).
This higher order convergence supports our explanation for the relatively lower convergence order found in the evolution of constraint violation (cf. the discussion in Sec.~\ref{sec:test}): indeed, the local hydrodynamic quantity follows higher order convergence if it is not close to the surface.

\begin{figure}
	\begin{center}
		\includegraphics[height=70mm]{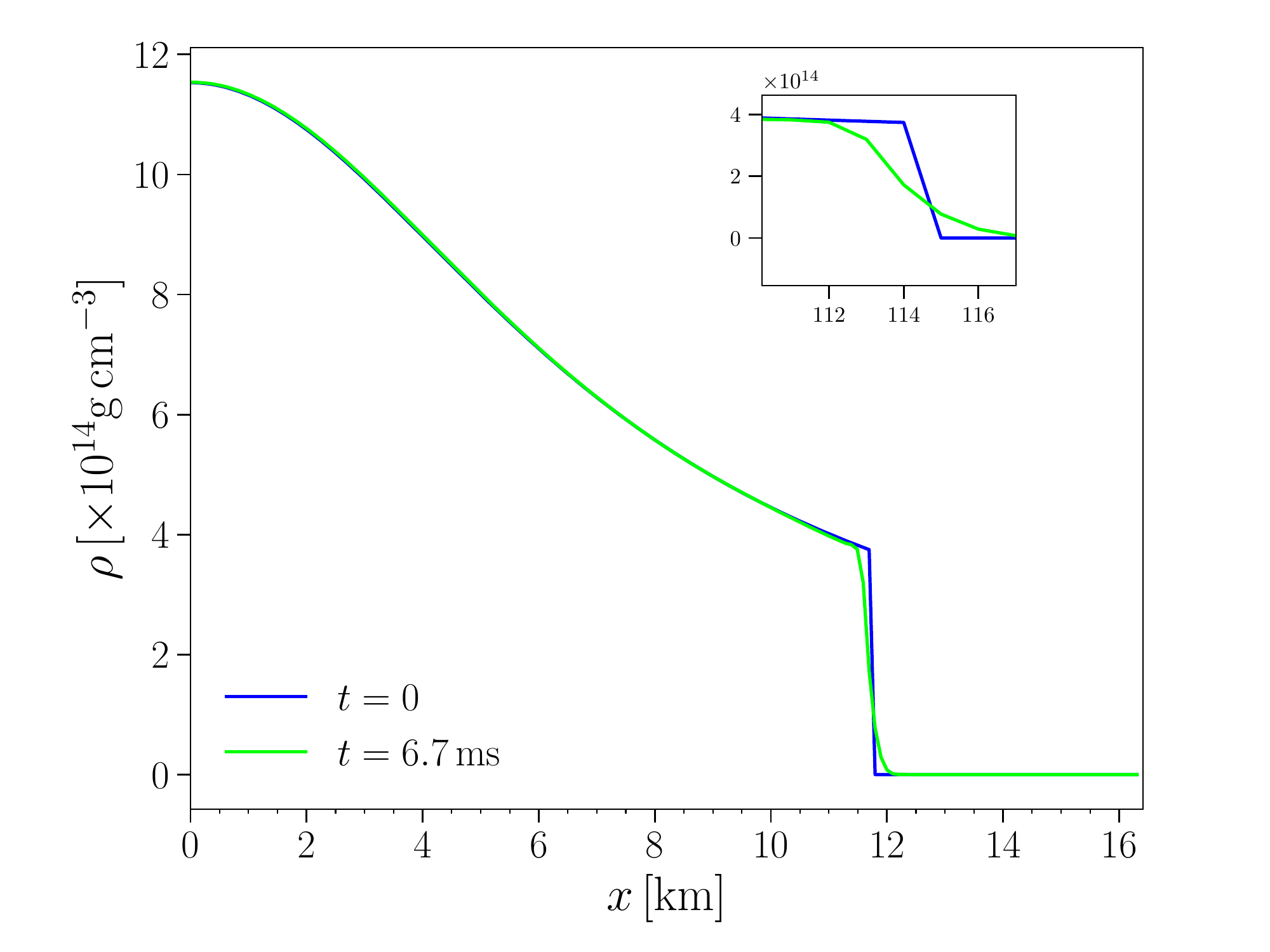}
	\end{center}
	\caption{Density profile along $x$-axis as in the initial condition (blue curve) and at the end of the test run (green curve, $t\sim6.7\,$ms) for the highest resolution case ($N=160$). The inset zooms in the region close to the surface. The $x$-axis of the inset indicates the index of the grid points in the finest FMR domain.}
	\label{fig:profile}
\end{figure}

Apart from local quantities, we have also investigated global density profile in radial direction, which could quantitatively demonstrate the capability of the implementation in resolving the discontinuous density across the surface of a QS.
The comparison of the initial and final (at $t\sim6.7\,$ms) density profile in the $N=160$ resolution case is shown in Fig.~\ref{fig:profile}. As can be seen, the density profile inside the QS is approximately unchanged during the entire evolution. The most obvious difference is the density discontinuity across the surface as can be seen from the inset in Fig.~\ref{fig:profile}: the very steep density drop at the surface becomes smoother during the evolution. Nevertheless, the density drop from $\rho_\mathrm{s}$ to 0 is resolved by $\sim$3 grid points in the end, which is similar to what can be seen from the rotating solution cases. 

Besides local quantities as mentioned above, we have also examined the performance of our method by looking at global quantities such as the conservation of baryonic mass and mass that becomes unbound during the simulation. Throughout the simulation, the change in baryonic mass stays as low as $\sim10^{-8}\,M_\odot$. The unbound matter is at a negligible amount of $\sim10^{-12}\,M_\odot$ even at its maximum, which is as expected for a static solution.


\section{On the Artificial Mass Ejection}
\label{sec:appeje}

In Sec.~\ref{sec:drot}, it is mentioned that unphysical mass ejection of $\sim10^{-4}\,M_\odot$ is found in the very beginning of the simulation for model MIT275dr. A possible explanation for this artificial mass ejection is that when the initial data is interpolated to the \SACRA grids for evolution, it is unavoidable that some part of the grids close to the surface of the QS might obtain a density which is finite but smaller than $\rho_\mathrm{s}$, as it is not possible for the Cartesian coordinates used in \SACRA to cover exactly the surface of the QS. 
And hence this part of the matter will be treated as atmosphere and might be driven off equilibrium in the later evolution. Given certain amount of initial energy perturbation due to the reason above, it should always be easier to eject matter from the surface of a QS if the angular velocity is larger as the binding energy for those matter are smaller.

To verify this explanation, and more importantly to know the reliability of the current implementation in estimating the amount of ejecta for binary merger case in the future, we have prepared several uniformly rotating QS models with different angular velocity and investigated the mass ejection of those models in the early stage of the evolution. We have constructed 3 models with $\rho_\mathrm{c}=1.152\times10^{15}\,\mathrm{g\,cm^{-3}}$ which is the same as the TOV solution we tested in previous section. The corresponding gravitational mass and angular velocity of those models are: $(M=2.15\,M_\odot,\Omega=4092.2\,\mathrm{rad\,s^{-1}})$, $(M=2.25\,M_\odot,\Omega=5768.0\,\mathrm{rad\,s^{-1}})$ and
$(M=2.37\,M_\odot,\Omega=7010.0\,\mathrm{rad\,s^{-1}})$. In addition, in order to verify that it is the angular velocity rather than mass that determines the final amount of the spurious mass ejection, we have constructed one additional model with smaller central density ($\rho_\mathrm{c}=6.304\times10^{14}\,\mathrm{g\,cm^{-3}}$) and with mass and angular velocity of $(M=2.25\,M_\odot,\Omega=7402.0\,\mathrm{rad\,s^{-1}})$. Furthermore, in order to estimate the maximum possible spurious mass ejection, we have constructed a model with mass and angular velocity very close to the mass shedding limit ($M=2.95\,M_\odot,\Omega=8682.4\,\mathrm{rad\,s^{-1}}$)\footnote{The uniform-rotation mass shedding limit of this EoS model is found at $M=3.02\,M_\odot$ and $\Omega=9149.6\,\mathrm{rad\,s^{-1}}$.}.

\begin{figure}
	\begin{center}
		\includegraphics[height=70mm]{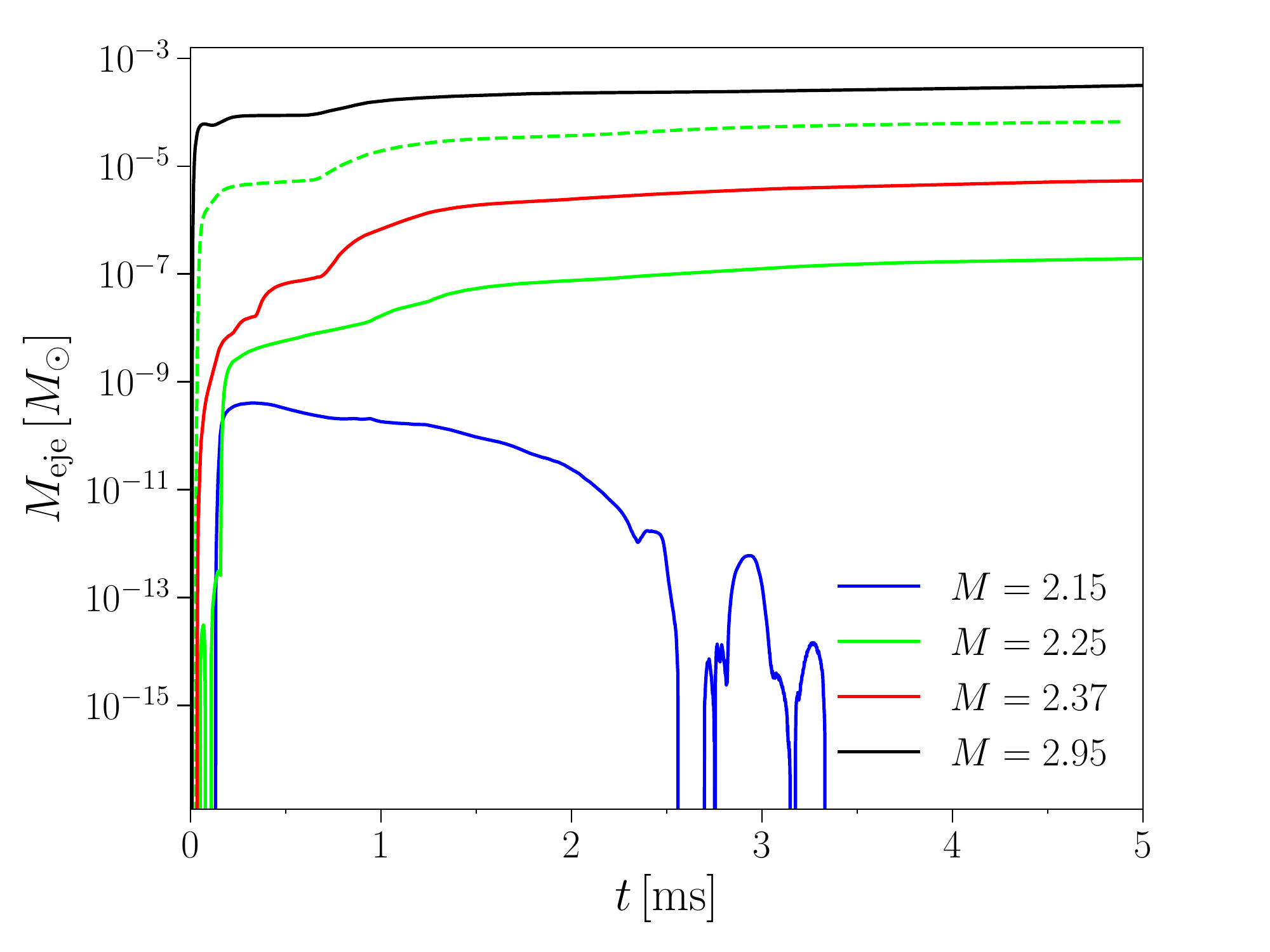}
	\end{center}
	\caption{The amount of unbound mass (according to the criterion that $u_t<-1.0$) for the four uniformly rotating QS models described in Appdendix~\ref{sec:appeje} in the $N=160$ resolution case. The solid red, green and blue curves indicate the results of the three models with central density of $1.152\times10^{15}\,\mathrm{g\,cm^{-3}}$ and the dashed green curve is the model with central density of $6.304\times10^{14}\,\mathrm{g\,cm^{-3}}$, which possesses the largest angular velocity although its mass is not the largest. The solid black curve shows the result of the model close to the uniform-rotation mass shedding limit. It is found that the artificial mass ejection is strongly dependent on the angular velocity.}
	\label{fig:ejeome}
\end{figure}

The result is shown in Fig.~\ref{fig:ejeome}. The artificial mass ejection monotonically increases with the angular velocity. Particularly, the comparison between the dashed green curve and the other curves clearly show that the amount of artificial ejection is affected mostly by the angular velocity of the model instead of the mass. The dependence still holds even if we consider the TOV solution case as reported in Appdendix~\ref{sec:apptov} and the MIT275dr model. 
As a consequence, the systematic uncertainty in the quantitative measurement of physical ejection will be higher for models with larger angular velocity. A maximum uncertainty will be order of $10^{-4}\,M_\odot$ as indicated by the results of the model $M=2.95\,M_\odot$.

\begin{figure}
	\begin{center}
		\includegraphics[height=70mm]{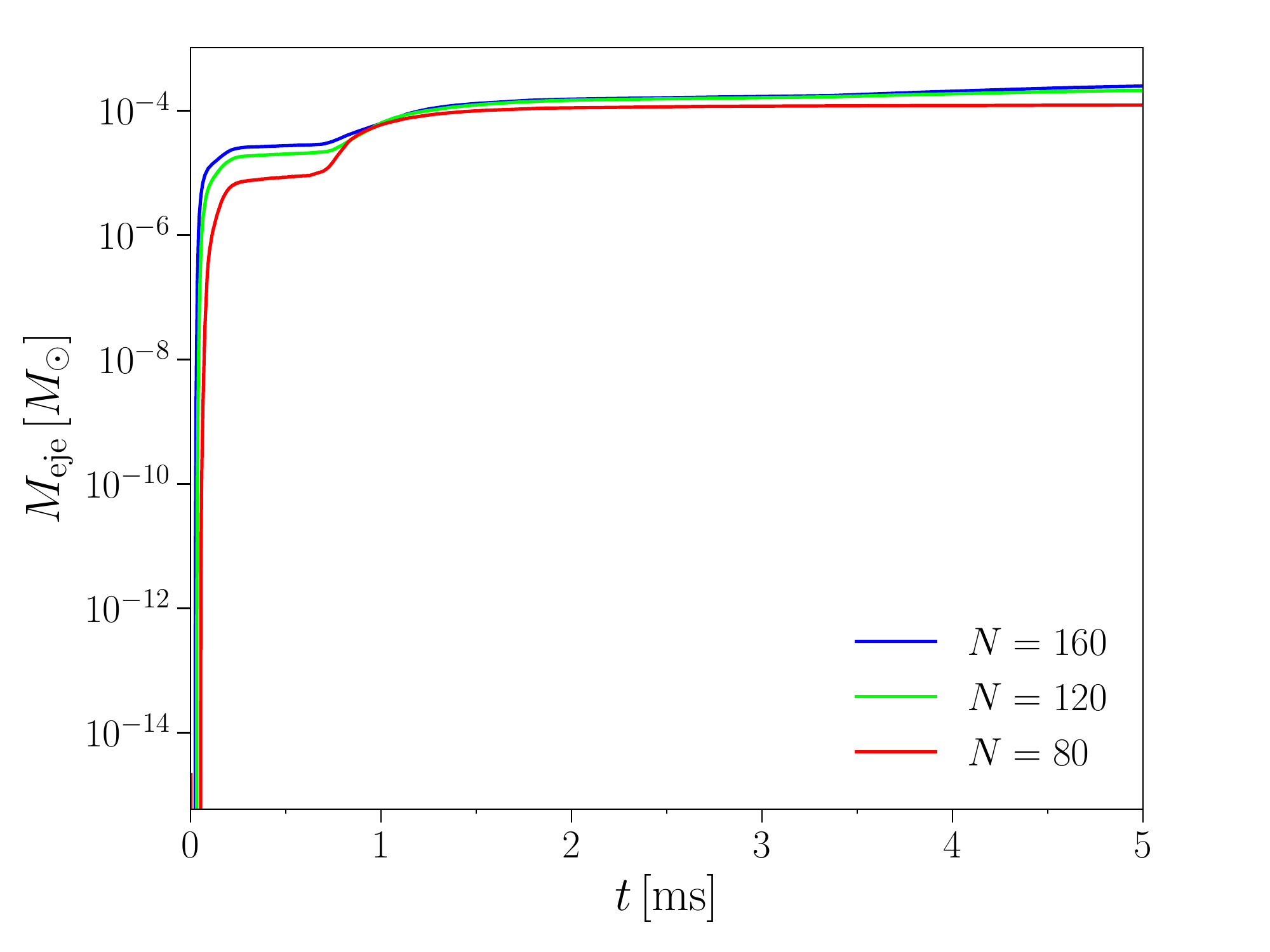}
	\end{center}
	\caption{The amount of unbound mass (according to the criterion that $u_t<-1.0$) for model MIT275dr in three different resolutions.}
	\label{fig:ejeconv}
\end{figure}

In addition, since the initial artificial mass ejection is related to the surface part of the star where both density and pressure gradient are discontinuous, the convergence behavior is expected to be worse than other part of the star. We have done the test on the model MIT275dr in three different resolutions and the result is shown in Fig.~\ref{fig:ejeconv}. As can be seen, the influence of the resolution on the amount of ejecta is very weak. Especially, the result is approximately the same for $N=120$ and $N=160$ cases. Further increasing the resolution within the current affordability of computational resources does not seem to be a viable way to resolve the problem.

For the purpose of extracting the amount of ejecta for binary quark star mergers, the relevant angular velocity should be order of $\sim2000\,\mathrm{rad\,s^{-1}}$ for a typical $1.4\,M_\odot$-$1.4\,M_\odot$ binary at a separation of $40\,$km, for which the initial artificial mass ejection should be much less than the MIT275dr model considered in this paper. Our preliminary BQS merger simulation confirmed this and we will cover more details about the binary case in the future.

\bibliographystyle{apsrev4-1}
\bibliography{aeireferences}

\end{document}